\pdfoutput=1

\documentclass[11pt,twoside,a4paper,cmspaper,final,collab]{cms-tdr}
\def\svnVersion{426227P}\def\cmsCernNoTag{CERN-EP-2017-035}\def\cmsCernDate{\today}\def\cmsMessage{Published in the Journal of High Energy Physics as \href{http://dx.doi.org/10.1007/JHEP09(2017)053}{\doi{10.1007/JHEP09(2017)053.}}}

\begin{document}\cmsNoteHeader{B2G-16-013}

\hyphenation{had-ron-i-za-tion}
\hyphenation{cal-or-i-me-ter}
\hyphenation{de-vices}
\RCS$Revision: 426156 $
\RCS$HeadURL: svn+ssh://svn.cern.ch/reps/tdr2/papers/B2G-16-013/trunk/B2G-16-013.tex $
\RCS$Id: B2G-16-013.tex 426156 2017-09-18 15:43:21Z eusai $
\newlength\cmsFigWidth
\ifthenelse{\boolean{cms@external}}{\setlength\cmsFigWidth{0.85\columnwidth}}{\setlength\cmsFigWidth{0.4\textwidth}}
\ifthenelse{\boolean{cms@external}}{\providecommand{\cmsLeft}{top\xspace}}{\providecommand{\cmsLeft}{left\xspace}}
\ifthenelse{\boolean{cms@external}}{\providecommand{\cmsRight}{bottom\xspace}}{\providecommand{\cmsRight}{right\xspace}}
\providecommand{\PQT}{\ensuremath{\mathrm{T}}\xspace}

\cmsNoteHeader{B2G-16-013}
\title{Search for a heavy resonance decaying to a top quark and a vector-like top quark at $\sqrt{s} = 13\TeV$}

\date{\today}

\abstract{
A search is presented for massive spin-1 $\PZpr$ resonances decaying to a top quark and a heavy vector-like top quark partner T. The search is based on a 2.6\fbinv sample of proton-proton collisions at 13\TeV collected with the CMS detector at the LHC. The analysis is optimized for final states in which the T quark decays to a W boson and a bottom quark. The focus is on all-jet final states in which both the W boson and the top quark decay into quarks that evolve into jets. The decay products of the top quark and of the W boson are assumed to be highly Lorentz-boosted and cannot be reconstructed as separate jets, but are instead reconstructed as merged, wide jets. Techniques for the identification of jet substructure and jet flavour are used to distinguish signal from background events.
Several models for $\PZpr$ bosons decaying to T quarks are excluded at 95\% confidence level, with upper limits on the cross section ranging from 0.13 to 10\unit{pb}, depending on the chosen hypotheses. This is the first search for a neutral spin-1 heavy resonance decaying to a top quark and a vector-like T quark in the all-hadronic final state.
}

\hypersetup{%
pdfauthor={CMS Collaboration},%
pdftitle={Search for a heavy resonance decaying to a top quark and a vector-like top quark at sqrt(s) = 13 TeV},%
pdfsubject={CMS},%
pdfkeywords={CMS, physics, heavy boson, vector-like quarks}}

\maketitle

\section{Introduction}
\label{introduction}
Many theoretical models of physics beyond the standard model (SM) predict the existence of heavy bosonic resonances \cite{mssm,nsd,nsd2,nsd3,nsd4,littlehiggs,ed,rs1,rs2}. Such resonances include $\PZpr$ gauge bosons \cite{leptophobicZprime, Zprimettxs, ZprimeCoupledtoGen3} and Kaluza--Klein excitations of a gluon in Randall--Sundrum models \cite{WarpedGaugeBosons, ExtraDim}. In many cases the couplings of these resonances to third-generation SM quarks are enhanced, leading to decay channels containing top quarks.

The CMS and ATLAS Collaborations at the CERN LHC have performed several searches for heavy resonances decaying to top quark-antiquark pairs (\ttbar) \cite{7tevZprime_CMSAllHad, 7tevZprime_CMSSemilept, 7tevZprime_ATLASAllHad, 7tevZprime_ATLASSemilept, 8tevZprime_CMSAllHadSemilept, 8tevZprime_CMSAllHadSemileptLept, 8tevZprime_ATLASSemilept}, placing very stringent limits on their production cross sections in the accessible kinematic range.
However, in models with a heavy gluon \cite{Bini2012,1126-6708-2009-06-001}, a composite Higgs boson \cite{Greco2014}, or extra spatial dimensions \cite{Bini2012,PhysRevD.89.095027}, an additional fermionic sector may be present in the form of a nonchiral (or vector-like) fourth generation of quarks. Topologies in which the $\PZpr$ boson decays into vector-like quarks have not yet been investigated experimentally. This search focuses on the kinematic range in which $\PZpr$ boson decays to tT dominate over those to TT, where T is a vector-like heavy quark with a charge of two thirds.

Vector-like quarks are fermions whose left- and right-handed components transform in the same way under the electroweak symmetry group of the SM. Consequently, their masses can be generated through direct
mass terms in the Lagrangian,  rather than via Yukawa couplings. This feature makes theories that include a heavy vector-like quark sector compatible with current Higgs boson measurements \cite{Khachatryan:2016vau}.

We present results of the first search for neutral spin-1 heavy resonances decaying to a top quark and a vector-like quark, in all-jet final states.  The search utilizes data from proton-proton collisions at a centre-of-mass energy of 13\TeV.  The analysis is optimized for the $\PQT\to\PQb\PW$ decay mode, but also considers the $\PQT \to \PQt\PH$ and $\PQT \to \PQT\Z$ decays.

The results of the analysis are compared with the predictions of two theoretical models.
The first model \cite{Bini2012} is an effective theory with one warped extra dimension that considers only the lowest-energy spin-1 and spin-1/2 resonances to describe the decays of the lightest Kaluza--Klein excitation of the gluon, $\mathrm{G}^*$, to one SM particle and one heavy fermion.
We consider the specific case where the $\mathrm{G}^*$ resonance decays to a top quark and a heavy top quark partner T. The model assumes branching fractions ($\mathcal{B}$) to be 50/25/25\% for T quark decay to the bW/tH/tZ channels. Benchmark values of $\tan{\theta_3}=0.44$, $\sin{\phi_{tR}}=0.6$, and $Y_* = 3$ are used for the model parameters. These benchmark values enhance the decays of the heavy resonance to a SM quark and a vector-like quark. The definitions of the parameters, the choice of their values, and their impact on the cross section are explained in Ref. \cite{Bini2012}, and the significant discovery potential at the LHC even with a comparatively small integrated luminosity is discussed. This model predicts the existence of other vector-like quarks such as the $\mathrm{T}_{5/3}$ quark, with a charge of five-thirds and a mass lower than the mass of the T quark. These other heavy quarks can have a moderate impact on the branching fraction of the $\mathrm{G}^*$ resonance to Tt, and their contribution is properly taken into account when comparing the model with the results of the analysis. In particular, the $\mathrm{T}_{5/3}$ quark becomes relevant when the mass of the $\mathrm{G}^*$ resonance is twice its mass.

The second model \cite{Greco2014} is a minimal composite effective theory of the Higgs boson based on the coset $SO(5)/SO(4)$, describing the phenomenology of heavy vector resonances, with particular focus on their interactions with top quark partners. The results of the analysis are compared with the cross section for the production of a neutral spin-1 resonance $\rho^{0}_{L}$ decaying to a top quark and a heavy top quark partner T. The model assumes T branching fractions to tH/tZ channels of 50/50\%. The following are benchmark values of the model parameters: $y_{L}=c_3=c_2=1$, and $g_{\rho_{L}}=3$. The model parameters and the choice of benchmark values are described in \cite{Greco2014}. This model is used to simulate signal samples.

\begin{figure}[htbp]
\centering
\includegraphics[width=0.50\textwidth]{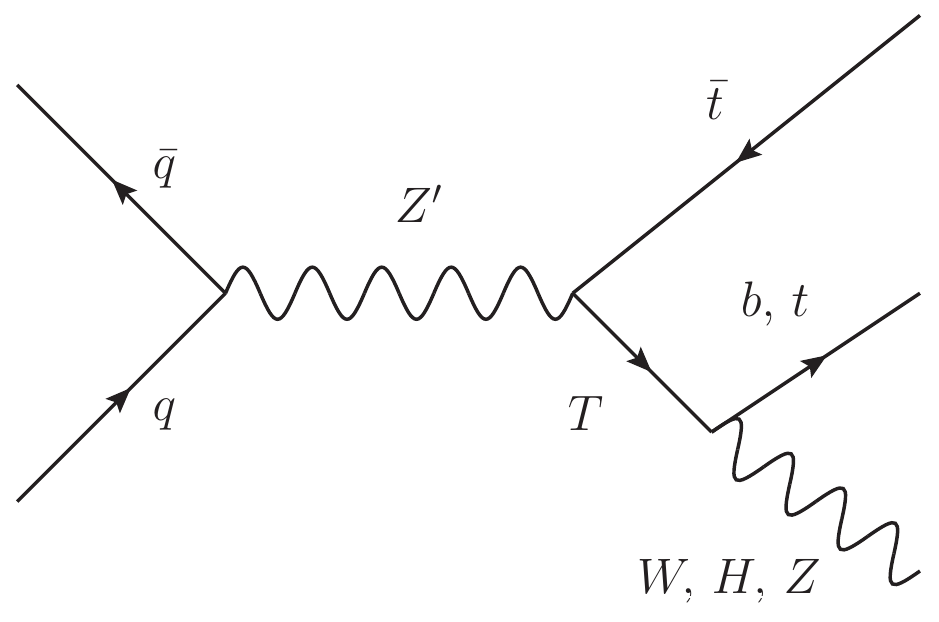}
\caption{The leading order Feynman diagram showing the production mode of the $\PZpr$ boson and its decay chain.}
\label{feynman}
\end{figure}

The $\mathrm{G}^*$ and the $\rho^{0}_{L}$ resonances are candidates for the $\PZpr$ of this search and are both produced through quark-antiquark pair interactions at the LHC. The kinematic distributions of the decay modes considered are comparable between the two models.
Hypothetical top quark flavour-changing neutral currents generated in the interaction between the top quark, $\PZpr$ boson, and T quark are estimated to be below the reach of current measurements \cite{Olive:2016xmw} because of the large suppression generated by off-shell effects of the $\PZpr$ boson and the T quark. The leading order Feynman diagram for the production of the $\PZpr$ boson and the decay chain under consideration is depicted in Fig.~\ref{feynman}.

Because of the large difference in mass between the W boson and the T quark, the W boson receives a large Lorentz boost, such that its decay products appear as merged jets (in a highly-boosted topology). Jet substructure algorithms are employed to reconstruct and identify the W boson originating from the decay of the T quark. If the mass difference between the $\PZpr$ boson and the T quark is much larger than the mass of the top quark, the top quark from the decay of the $\PZpr$ boson also receives a large transverse momentum (\pt), in which case jet-substructure techniques can also be used to identify and reconstruct the all-jets decay of the top quark.

The dominant background is from SM events and is comprised of jets produced through the strong interaction, i.e. quantum chromodynamics (QCD) multijet events, followed by events from \ttbar pair production and from single top quark production.
The contribution of the latter processes is estimated from simulation, while the multijet QCD background is estimated from data using signal-depleted control regions.

This paper is organized as follows: Section~\ref{det} gives a description of the CMS detector and the reconstruction of events. Section \ref{samples} describes the data and the simulated samples used in the analysis. An overview of the jet-substructure algorithms and the details of the selection for the analysis are given in Section~\ref{recosel}.  Estimation of SM background processes is discussed in Section~\ref{background}, while Section~\ref{systematic} describes the systematic uncertainties. The results of the analysis and a summary are given in Sections \ref{results} and \ref{conclusions}, respectively.

\section{The CMS detector}
\label{det}

The central feature of the CMS apparatus is a superconducting solenoid of 6\unit{m} internal diameter, providing a magnetic field of 3.8\unit{T}. Within the solenoid volume are a silicon pixel and strip tracker, a lead tungstate crystal electromagnetic calorimeter (ECAL), and a brass and scintillator hadron calorimeter (HCAL), each composed of a barrel and two endcap sections. Forward calorimeters extend the pseudorapidity ($\eta$) coverage provided by the barrel and endcap detectors. Muons are measured in gas-ionization detectors embedded in the steel flux-return yoke outside the solenoid.

A particle-flow event algorithm~\cite{CMS-PAS-PFT-09-001,CMS-PAS-PFT-10-001} reconstructs and identifies each individual particle with an optimized combination of information from the various elements of the CMS detector. The energy of photons is directly obtained from the ECAL measurement, corrected for zero-suppression effects. The energy of electrons is determined from a combination of the electron momentum at the primary interaction vertex as determined by the tracker, the energy of the corresponding ECAL cluster, and the energy sum of all bremsstrahlung photons spatially compatible with originating from the electron track. The energy of muons is obtained from the curvature of the corresponding track. The energy of charged hadrons is determined from a combination of their momentum measured in the tracker and the matching ECAL and HCAL energy deposits, corrected for zero-suppression effects and for the response function of the calorimeters to hadronic showers. Finally, the energy of neutral hadrons is obtained from the corresponding corrected sum of ECAL and HCAL energies. Primary vertices are reconstructed
using a deterministic annealing filter algorithm~\cite{Chatrchyan:2014fea}.
The vertex with the largest sum of the squares of the associated track $\pt$ values
is taken to be the primary event vertex. A more detailed description of the CMS detector, together with a definition of the coordinate system used and the relevant kinematic variables, can be found in Ref.~\cite{Chatrchyan:2008zzk}.

\section{Data and simulation samples}
\label{samples}

The analysis is based on data from proton-proton collisions collected in 2015 by the CMS experiment at a centre-of-mass energy of 13\TeV, corresponding to a total integrated luminosity of 2.6\fbinv.
The events are selected with an online trigger that required the scalar \pt sum of the jets (\HT) to be larger than 800\GeV. The offline  \HT  is required to be larger than 850\GeV. After this selection, the trigger is more than 97\% efficient in selecting those events that would pass the analysis selection. The trigger and offline  \HT  selections do not significantly impact the overall signal efficiency because the masses of the spin-1 resonances considered in this analysis are at least 1.5\TeV.

The signal processes are simulated using \MADGRAPH~v5.2.2.2~\cite{Alwall:2014hca}. Neutral spin-1 resonances ($\PZpr$ boson) decaying exclusively to a top quark and an up-type heavy vector-like quark (T) are generated.
Data samples are produced for three values of mass of the $\PZpr$ boson and a width of 1\% the mass. For the T quark samples the width of the quark is fixed to 1\MeV. The values of the width are chosen to be much smaller than the detector resolution. The T quark is generated with left-handed chirality. The impact of the chirality of the T quark on the analysis is assessed on a single signal configuration and is found to be insignificant, and for this reason the right-handed chirality case is not explicitly considered.

The simulation of the signal event production is based on a simplified low-energy effective theory describing the phenomenology of heavy vector resonances in the minimal composite Higgs model \cite{Greco2014}.
Signal samples are generated for three decay modes of the T quark: $\PQT\to\PQb\PW$, $\PQt\PH$, and $\PQt\Z$.
Several mass hypotheses for the $\PZpr$ (T) resonance are considered ranging from 1.5 to 2.5 (0.7 to 1.5)\TeV. The combination of the $\PZpr$ and T masses is chosen such that the mass of the T quark is roughly 1/2, 2/3, or 5/6 of the $\PZpr$ boson mass.  For some of the samples generated, the top quark from the decay of the $\PZpr$ boson receives a small \pt and its decay does not result in a boosted topology.

The decay of heavy resonances in signal events is processed with \textsc{MadSpin} \cite{Artoisenet2013} to correctly treat the spin correlations in the decay chain.
The matrix element calculations for signal processes include one extra parton at most emitted at tree level.
To model fragmentation and parton showering, the \PYTHIA~8.2~\cite{Sjostrand:2014zea} tune CUETP8M1 \cite{pythiatune} is used, and the MLM scheme~\cite{mlm} is used to match parton emission in the matrix element with the parton shower. Differential jet rates are checked for smoothness to ensure that the matching scale is chosen correctly.

Background top quark pair production is simulated with the next-to-leading-order generator
\POWHEG~\textsc{v2} \cite{Nason:2004rx,Frixione:2007vw,Alioli:2010xd,Frixione:2007nw,Re:2010bp}.
The \ttbar event sample is normalized to the next-to-next-to-leading order (NNLO) cross section of $\sigma_{\ttbar} = 831.76\unit{pb}$~\cite{Czakon:2011xx}.
Background events from single top quark production in the tW channel are also generated with \POWHEG~\textsc{v2} and are normalized to a cross section of 71.7\unit{pb}~\cite{Kidonakis:2013zqa}. Single top quark production in the $s$ and $t$ channels without an associated W boson is generated with \MADGRAPH~v5.2.2.2~\cite{Alwall:2014hca} and the cross sections are normalized to 10.32 and 216.99\unit{pb}, respectively~\cite{Aliev:2010zk,Kant:2014oha}. All samples are interfaced to \PYTHIA~8.2 for fragmentation and parton showering.
The multijet QCD production is estimated from data. Simulated multijet QCD events are used only to validate the method of background estimation and are generated with \PYTHIA~8.2, binned in
\HT to increase  the event sample in the high-energy region.

All events were generated with the NNPDF~3.0 parton distribution functions (PDFs)~\cite{Ball:2014uwa}.
All simulated event samples include the simulation of additional inelastic proton-proton interactions
within the same or adjacent bunch crossings (pileup). The detector response is simulated with the \GEANTfour package~\cite{Agostinelli:2002hh,Allison:2006ve}. Simulated events are processed through the same software chain as used for collision data and are reweighted to match the observed distribution of the number of pileup interactions in data.

\section{Event reconstruction and selection}
\label{recosel}
For each event, hadronic jets are clustered from the reconstructed particles with the infrared and collinear safe anti-\kt algorithm ~\cite{ktalg}, using the \FASTJET~3.0 software package~\cite{fastjet1,Cacciari:2011ma} with the distance parameters $R=0.4$ (AK4 jets) and $0.8$ (AK8 jets). The two types of jets are reconstructed independently.
Charged hadrons not associated with the primary vertex of the interaction are not considered when clustering.
Corrections based on the jet area~\cite{jetarea_fastjet_pu} are applied to remove the energy contribution
of neutral hadrons arising from pileup collisions.
Further corrections are used to account
for the nonlinear calorimeter response
as a function of $\eta$ and $\pt$~\cite{Chatrchyan:2011ds}, derived from simulation and from data-to-simulation correction factors. Spurious jets due to detector noise effects are removed by requiring that neutral particles contribute less than 99\% of the electromagnetic and hadronic energy in a jet.
Only jets with $\abs{\eta}<2.4$ are considered; no requirements on lepton or imbalance in transverse momentum are applied.

This analysis considers signal events characterized by a three-jet topology. One of the jets corresponds to the boosted top quark from the decay of the $\PZpr$ boson, the second originates from the W boson  of the T quark decay, and the third is from the b quark emitted in the T quark decay. These selection criteria are optimized for the decay of the T quark to bW, but the analysis is sensitive to the other decay modes of the T quark as well.
To identify $\cPqt$~jets, the jets associated with top quarks, the ``CMS top tagger v2''~\cite{CMS-PAS-JME-15-002} algorithm
is used. In this algorithm, the constituents of the AK8 jets are reclustered using the
Cambridge--Aachen algorithm~\cite{CAcambridge,CAaachen}. The
modified mass-drop tagger algorithm~\cite{mmdt}, also
known as the ``soft drop'' algorithm with angular exponent $\beta = 0$,
soft threshold $z_\text{cut} < 0.1$,
and characteristic radius $R_{0} = 0.8$~\cite{softdrop},
is used to remove soft, wide-angle radiation from the jet.
This algorithm identifies two subjets within the AK8 jet corresponding to the b jet and the decay of the W boson.
Additionally, the ``N-subjettiness'' variables $\tau_\mathrm{N}$~\cite{Thaler:2010tr, Thaler:2011gf} are used. These variables, calculated using all the particle-flow constituents of the AK8 jet, quantify the degree to which a jet can be regarded as composed of N subjets.

For the identification of top quark candidates, the soft-drop mass, $m_{\text{SD}}$, is required to satisfy $110<m_{\text{SD}}<210\GeV$ and the N-subjettiness variable is required to satisfy $\tau_{3}/\tau_{2}<0.86$. These selections correspond to a misidentification rate of 10\% for multijet QCD, and an efficiency greater than 70\%.
To ensure that the decays of the top quark are merged in a single jet,  AK8 jets are required to have $\pt > 400\GeV$.
Jets satisfying the aforementioned momentum, mass, and N-subjettiness selections are referred to as
``$\cPqt$-tagged''.

For the identification of W jets, the same jet reclustering procedure as in the t tagging algorithm is chosen. Additionally, jets are required to fulfill $70<m_{\text{SD}}<100\GeV$, $\tau_{2}/\tau_{1}<0.6$, and $\pt>200$\GeV. These criteria correspond to a misidentification rate of approximately 5\% for multijet QCD, and an efficiency of approximately 60\% for genuine W bosons not coming from the decay of a top quark. Jets satisfying these requirements are referred to as ``W-tagged''.

The Combined Secondary Vertex v2 (CSVv2) algorithm~\cite{CMS-PAS-BTV-15-001,Chatrchyan:2012jua} is used to identify AK4 jets originating from b quarks (b tagging).  The `medium' working point of the algorithm is used, which provides an efficiency of approximately 70\% for the identification of genuine b quark jets while rejecting 99\% of light-flavour jets. The `loose' working point of the algorithm is used for the background estimation, providing an efficiency of approximately 85\% and a light-flavour rejection rate of 90\%. Additionally, $\cPqt$-tagged jets with a b-tagged subjet~\cite{Chatrchyan:2012jua,8tevZprime_CMSAllHadSemileptLept} are used to improve the discrimination power against background processes.
The CSVv2 algorithm with the `medium' working point is used for subjet b tagging.

The events are required to have at least one b-tagged AK4 jet~\cite{CMS-PAS-BTV-15-001}, with $\pt>100\GeV$ and $\abs{\eta}<2.4$. To avoid possible overlaps, the AK4 jet is required to have an angular separation, $\Delta R$, of at least 0.8 with respect to the t-tagged jet and the W-tagged jet. The angular separation variable $\Delta R$ is defined as $\sqrt{\smash[b]{(\Delta\phi)^2+(\Delta\eta)^2}}$, where $\phi$ is the azimuthal angle.
Among the b jets satisfying these requirements, the one with the highest \pt is selected.
The T quark candidate four-momentum is defined as the sum of the 4-vectors of the selected b jet and the W-tagged jet. Only events with a T quark candidate  mass $m_{\PQT}>500$\GeV are considered. This selection criterion helps to reject the  \ttbar background.
The reconstructed $\PZpr$ boson candidate four-momentum is defined as the sum of the 4-vectors of the T quark candidate and the selected t-tagged jet. The invariant mass of the $\PZpr$ boson candidate $m_{\PZpr}$ is used as the main discriminating observable in the analysis.

Events are grouped into two separate categories according to the presence or absence of a b-tagged subjet in the t-tagged jet. Events containing a b-tagged subjet are placed in the ``SR 2 b tag category'' as they contain one b-tagged AK4 jet together with a b-tagged subjet associated with the t-tagged jet, as opposed to events in the ``SR 1 b tag category'' that contain only one b-tagged AK4 jet. No selection criteria are applied to specifically target the tH and tZ final states of the T quark.

Table \ref{tab:signal_efficiency} shows the selection efficiency for the signal in the different event categories. The samples with the smallest difference in mass between the $\PZpr$ boson and T quark have a degraded reconstruction efficiency because of the low \pt of the top quark originating from the decay of the $\PZpr$ boson. For several mass points the reconstruction efficiency is higher for the T $\to$ tH or T $\to$ tZ decay channel than for T $\to$ bW, for which the analysis is optimized. This is because if the T quark decays to a t quark instead of a b quark, there are two t quarks in the final state, hence it is more likely that at least one of the two t quarks will be tagged. In addition to this, t quarks coming from the decay of a T quark have a higher \pt, therefore are more likely to be tagged.

\begin{table}[hbtp]
\centering
\topcaption{Selection efficiencies for the signal in the categories used in the analysis. The quoted uncertainties are statistical.}
\label{tab:signal_efficiency}
\begin{tabular}{cccc}
\hline
\multicolumn{4}{c}{$\mathcal{B}(\PQT\to \PQb\PW)=1$}\\
\hline
$m_{\PZpr}$ [\GeVns{}] & $m_{\PQT}$ [\GeVns{}] & Efficiency SR 1 b tag [\%] & Efficiency SR 2 b tag [\%]  \\
\hline
1500 & 700 &  $1.2 \pm 0.2$ & $1.9 \pm 0.3$ \\
1500 & 900 &  $0.74 \pm 0.17$ & $1.1 \pm 0.2$ \\
1500 & 1200 &  $0.23 \pm 0.09$ & $0.21 \pm 0.09$ \\
2000 & 900 &  $2.6 \pm 0.3$ & $3.6 \pm 0.4$ \\
2000 & 1200 &  $2.1 \pm 0.3$ & $3.0 \pm 0.4$ \\
2000 & 1500 &  $0.89 \pm 0.18$ & $0.87 \pm 0.18$ \\
2500 & 1200 &  $3.3 \pm 0.4$ & $3.9 \pm 0.4$ \\
2500 & 1500 &  $2.8 \pm 0.3$ & $3.6 \pm 0.4$ \\[1.5ex]
\hline
\multicolumn{4}{c}{$\mathrm{\mathcal{B}(T\to tH)}=1$}\\
\hline
$m_{\PZpr}$ [\GeVns{}] & $m_{\PQT}$ [\GeVns{}] & Efficiency SR 1 b tag [\%] & Efficiency SR 2 b tag [\%]  \\
\hline
1500 & 700 &  $0.55 \pm 0.15$ & $0.75 \pm 0.17$ \\
1500 & 900 &  $0.65 \pm 0.16$ & $0.93 \pm 0.19$ \\
1500 & 1200 &  $0.26 \pm 0.10$ & $0.37 \pm 0.12$ \\
2000 & 900 &  $1.8 \pm 0.3$ & $2.6 \pm 0.3$ \\
2000 & 1200 &  $2.0 \pm 0.3$ & $2.9 \pm 0.3$ \\
2000 & 1500 &  $1.7 \pm 0.3$ & $2.2 \pm 0.3$ \\
2500 & 1200 &  $2.9 \pm 0.3$ & $3.9 \pm 0.4$ \\
2500 & 1500 &  $3.0 \pm 0.3$ & $4.1 \pm 0.4$ \\[1.5ex]
\hline
\multicolumn{4}{c}{$\mathrm{\mathcal{B}(\PQT\to\PQt\Z)}=1$}\\
\hline
$m_{\PZpr}$ [\GeVns{}] & $m_{\PQT}$ [\GeVns{}] & Efficiency SR 1 b tag [\%] & Efficiency SR 2 b tag [\%]  \\
\hline
1500 & 700 &  $0.62 \pm 0.15$ & $0.84 \pm 0.18$ \\
1500 & 900 &  $0.78 \pm 0.17$ & $0.98 \pm 0.19$ \\
1500 & 1200 &  $0.50 \pm 0.14$ & $0.54 \pm 0.14$ \\
2000 & 900 &  $2.4 \pm 0.3$ & $3.1 \pm 0.4$ \\
2000 & 1200 &  $2.8 \pm 0.3$ & $3.9 \pm 0.4$ \\
2000 & 1500 &  $2.3 \pm 0.3$ & $2.8 \pm 0.3$ \\
2500 & 1200 &  $4.3 \pm 0.4$ & $5.4 \pm 0.5$ \\
2500 & 1500 &  $4.5 \pm 0.4$ & $6.0 \pm 0.5$ \\
\hline
\end{tabular}
\end{table}

\section{Background estimation}
\label{background}
There are two dominant source of background: multijet QCD production and  top quark production, including both \ttbar and single top quark contributions.  The multijet background contribution is the most important for this search. Approximately 20\% of the top quark production in the signal region is composed of single top quark events, mostly in the tW channel. Pair production of top quarks in association with a vector boson is not a relevant background for this analysis because of the non-boosted nature of the process and its relatively small cross section. Its contribution is estimated to be less than 0.3\% of the total number of events in the signal region.

The multijet background is derived from data with the following procedure. Sideband regions are defined by inverting the b tagging requirement on the AK4 jet for the selection of the signal. Specifically, the AK4 jet has to fail the b tagging requirement, using a `loose' operating point of the b tagging algorithm. Events with additional b-tagged jets are vetoed to ensure independence with respect to the signal region.
Two different sideband regions are used for the two signal categories according to the presence or absence of a b-tagged subjet in the t-tagged jet. A summary of the selection criteria is shown in Table~\ref{tab:selection}.

\begin{figure}[htbp]
\centering
\includegraphics[width=0.49\textwidth]{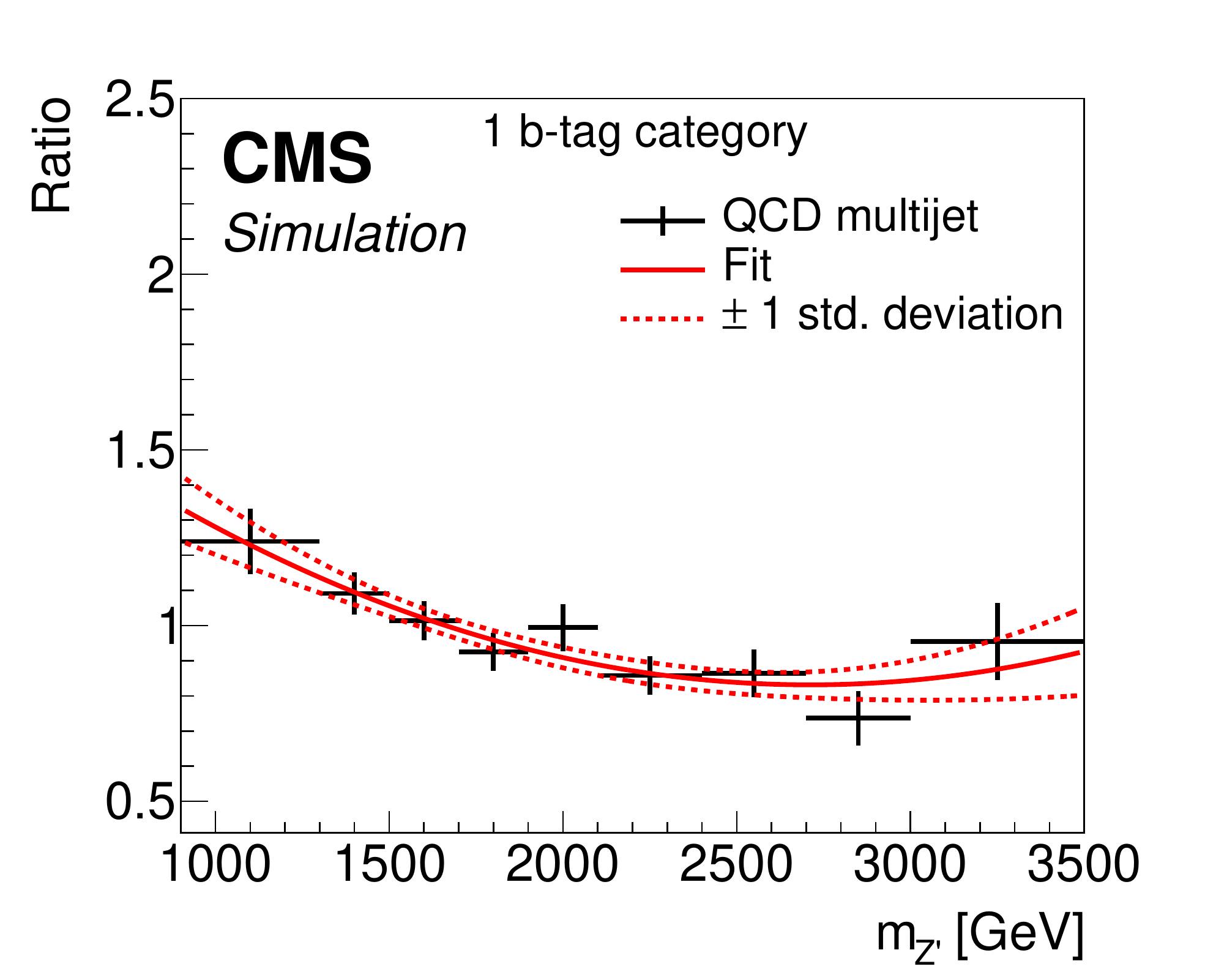}
\includegraphics[width=0.49\textwidth]{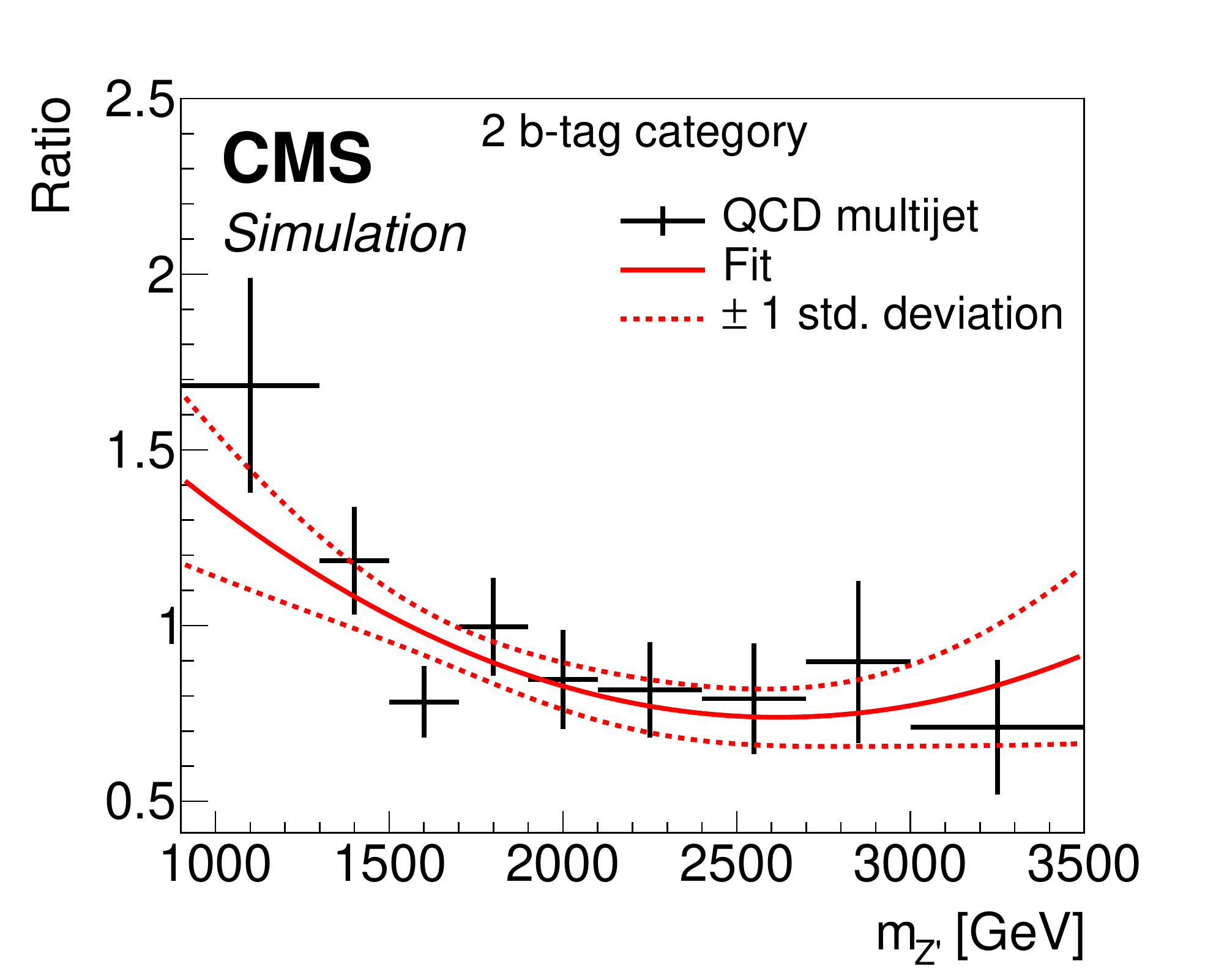}
\caption{Ratio of the number of events in the signal region to the number in the sideband region, as a function of the $\PZpr$ mass, for simulated background QCD multijet events. The left (right) plot involves events with no (at least one) b-tagged subjet. The solid line shows a fit of a second-order polynomial function to the ratio.}
\label{sidebandratio}
\end{figure}

The shape of the $m_{\PZpr}$ distribution is compared between the sideband region and the signal region in a sample of simulated multijet QCD events. Figure \ref{sidebandratio} shows the bin-by-bin ratio of the signal region to the sideband region.  Both histograms are normalized to unity before computing the ratio.

\begin{table}[hbtp]
\centering
\topcaption{Summary  of the selection criteria for the event categories in the signal region (SR) and the sideband region (SB).}
\label{tab:selection}
\begin{tabular}{lcccc}
\hline
Selection & SR 1 b tag & SB for 1 b tag & SR 2 b tag & SB for 2 b tag  \\
\hline
1 t tag and 1 W tag              & Yes & Yes & Yes & Yes \\
Subjet b tag on t-tagged jet     & Veto       & Veto & Yes       & Yes \\[0.5ex]
1 AK4 jet, $\pt>100$\GeV,& \multirow{2}{*}{Yes} &  \multirow{2}{*}{Yes} &  \multirow{2}{*}{Yes} &  \multirow{2}{*}{Yes} \\
$\Delta R(\PQt-/\PW-\text{jet}, \text{jet})>0.8$&&&&\\[0.5ex]
b tag on AK4 jet              & Yes & ``loose'' Veto & Yes & ``loose'' Veto \\
$m_{\PQT}>500$\GeV & Yes & Yes & Yes & Yes \\
\hline
\end{tabular}
\end{table}

The ratio is fit with a second-order polynomial function, which represents the correction factor required to weight the events in the sideband region to reproduce the shape of the multijet background in the signal region. This is the simplest functional form providing a satisfactory fit. To avoid double counting when estimating the multijet background from data, the  top quark contribution in the sideband region, estimated from simulation, is subtracted. Good agreement in shape between data and simulated events is observed in the sideband regions.

The normalization of the predicted multijet background cannot be reliably extracted from simulation and is fixed by a maximum likelihood fit to data in the signal region in a background-only hypothesis. The contribution from \ttbar is properly taken into account. A flat prior is used for the nuisance parameter associated with the normalization of the multijet background. The fit is performed on the $m_{\PZpr}$ distribution and, as a consistency check, on the $H_{\PQT}$ distribution, obtaining compatible results. It is verified that the scale factor obtained from the fit is not affected by changing the signal hypotheses considered in this analysis. The inclusive normalization factors are $0.093\pm0.004$ and $0.12\pm0.01$ for the 1 and 2 b tag event categories, respectively. This normalization is used for plots in Section~\ref{results}. For the extraction of upper cross section limits on signal production, the normalization of the multijet background is determined by the maximum likelihood fit to data described in Section~\ref{results}.

The top quark background is estimated using simulated event samples normalized to the theoretical cross sections, as listed in Section \ref{samples}. The systematic uncertainties that may impact the event rates and the shapes of the $m_{\PZpr}$ and $m_{\PQT}$ distributions in simulated events  are discussed in Section~\ref{systematic}.
Table \ref{tab:yields} shows the expected background yields for the two event categories, along with the observed number of events in data. The uncertainties  include both statistical and systematic components; the estimation of the latter is described in Section \ref{systematic}. The yields have been normalized to give the observed total numbers of events.

\begin{table}[hbtp]
\centering
\topcaption{Number of events in the two signal categories of the analysis. The uncertainties include both statistical and systematic components.}
\label{tab:yields}
\renewcommand*{\arraystretch}{1.4}
\begin{tabular}{lcc}
\hline
Sample & SR 1 b tag & SR 2 b tag  \\
\hline
QCD multijet & $1227^{+59}_{-59}$ & $222^{+22}_{-22}$\\
SM top quark & $81^{+31}_{-23}$ & $66^{+23}_{-18}$\\
\hline
Total background & $1308^{+67}_{-63}$ & $288^{+32}_{-29}$\\
Data & 1307 & 289\\
\hline
\end{tabular}
\end{table}

\section{Systematic uncertainties}
\label{systematic}

Several sources of systematic uncertainty may  impact the simulated signal and the top quark backgrounds.  The  procedure used to estimate the multijet background is subject to uncertainties as well. These systematic uncertainties affect both the shape and the normalization of the $m_{\PZpr}$ distribution used in the statistical procedure to infer the presence of signal. The systematic uncertainties are treated as nuisance parameters in the likelihood fit used to extract the upper cross section limit on signal production and are constrained by the data. Table~\ref{tab:syst_corr} reports the sources of systematic uncertainty, their  impact on event rates, the type (rate only, or rate and shape), and the processes for which they are relevant.

\begin{table}[hbtp]
\centering
\topcaption{Sources of systematic uncertainty, their impact on event rates, their  type, and the processes for which they are relevant.}
\label{tab:syst_corr}
\begin{tabular}{lcccc}
\hline
Systematic uncertainty & Rate 1 b tag & Rate 2 b tag & Type & Process\\
\hline
b-tagging efficiency & 9--14\% & 12--17\% & rate + shape & t bkg and signal\\
t-tagging efficiency & 8--14\% & 8--14\% & rate + shape & t bkg and signal\\
W-tagging efficiency & 0.1--6\% & 0.1--6\% & rate + shape & t bkg and signal\\
Jet energy scale & 0.4--10\% & 0.1--8\% & rate + shape & t bkg and signal\\
Jet energy resolution & 0--2\% &  0--2\% & rate + shape & t bkg and signal\\
Integrated luminosity  & 2.3\% & 2.3\% & rate & t bkg and signal\\
Trigger efficiency & 3\% & 3\% & rate & t bkg and signal\\
PDFs & 3--9\% & 3--8\% & rate + shape & t bkg and signal\\
Pileup reweighting & 0--3\% & 0.1--2	\% & rate + shape & t bkg and signal\\
$\mu_\mathrm{R}$, $\mu_\mathrm{F}$& 3--44\% & 1--41\%& rate + shape & t bkg and signal\\
Sideband corr. (fit unc.) & 4\% & 9\% & rate + shape & QCD multijet\\
Sideband corr. (fit form) & 1\% & 2\% & rate + shape & QCD multijet\\
Sideband norm. & 50\% & 50\% & rate & QCD multijet\\
\hline
\end{tabular}
\end{table}

The energy scale of jets~\cite{Chatrchyan:2011ds} is corrected with dedicated \pt- and $\eta$-dependent factors derived for AK4 and AK8 jets. The jet energy corrections for AK8 subjets are  the same as for  AK4 jets, scaled for the difference in jet area. Systematic uncertainties are derived by varying the jet energy scale within its uncertainty and thus obtaining the shape and normalization impact on the distribution of $m_{\PZpr}$.

The energy resolution of jets is lower in data than in simulation, and thus a smearing factor is applied to the four-vectors of AK4 jets, AK8 jets, and to the subjets, in simulated events. The smearing factor for subjets is the same as that for  AK4 jets.  The impact of this uncertainty, calculated by varying the smearing factor within its uncertainty, is negligible compared to that of the other uncertainties.

The discrepancy of the t tagging efficiency between data and simulation is corrected with scale factors derived in a semileptonic \ttbar topology using a ``tag-and-probe'' technique~\cite{CMS-PAS-JME-13-007,Khachatryan:2010xn}. This procedure selects a pure sample of \ttbar events using a tight selection on the leptonically decaying top quark. The sample is then used to measure the efficiency of the t tagging algorithm on the hadronically decaying top quark.  The scale factors are derived as a function of the jet \pt, along with their respective uncertainties.  A similar procedure is used to derive the correction factors for the W tagging algorithm. Jet and subjet b tagging efficiency correction factors for heavy- and light-flavour jets~\cite{CMS-PAS-BTV-15-001} are varied within their uncertainties to derive the impact on shape and normalization in simulated samples.

Different choices of the renormalization ($\mu_\mathrm{R}$) and factorization ($\mu_\mathrm{F}$) scales used to produce the simulated samples induce shape and normalization changes in the $\PZpr$ boson mass distribution.  The impact is assessed by using dedicated simulated top quark and signal events where the $\mu_\mathrm{R}$ and $\mu_\mathrm{F}$ are both scaled up or down by a factor of 2.

The pileup reweighting uncertainty is evaluated by varying the effective inelastic cross section by 5\%. To account for trigger efficiency discrepancies in data and simulation, a 3\% rate uncertainty is assigned to the simulated signal and top quark event yields. The uncertainty in the measurement of the integrated luminosity is calculated to be 2.3\% \cite{CMS-PAS-LUM-15-001}.

The systematic uncertainty related to the choice of the PDF values is assessed by varying the eigenvectors for the NNPDF 3.0 set used in the simulation. The variations are summed in quadrature to obtain the shape and rate variation due to PDF effects.

The systematic uncertainty in the estimation of the multijet background arises from the sideband shape correction function (weight function) as explained in Section \ref{background}. When fitting the ratio between the sideband and the signal region, the statistical uncertainties of the simulated samples in the procedure are considered. In addition, a linear functional form for the weight function is tested for comparison, and the observed difference is taken into account as a systematic uncertainty. These uncertainties are propagated through the background estimation procedure to obtain their impact on the shape and normalization of the $m_{\PZpr}$ distribution. The normalization of the multijet background is determined during the limit setting procedure by allowing it to vary within an uncertainty of 50\% in the maximum likelihood fit to data.

The most significant uncertainties are the ones associated with the multijet background fit function, and with the choice of renormalization and factorization scales. Assigning a 50\% uncertainty to the multijet background normalization does not significantly affect the results obtained in Section~\ref{results}.
\section{Results}
\label{results}
\begin{figure}[htbp]
\centering
\includegraphics[width=0.60\textwidth]{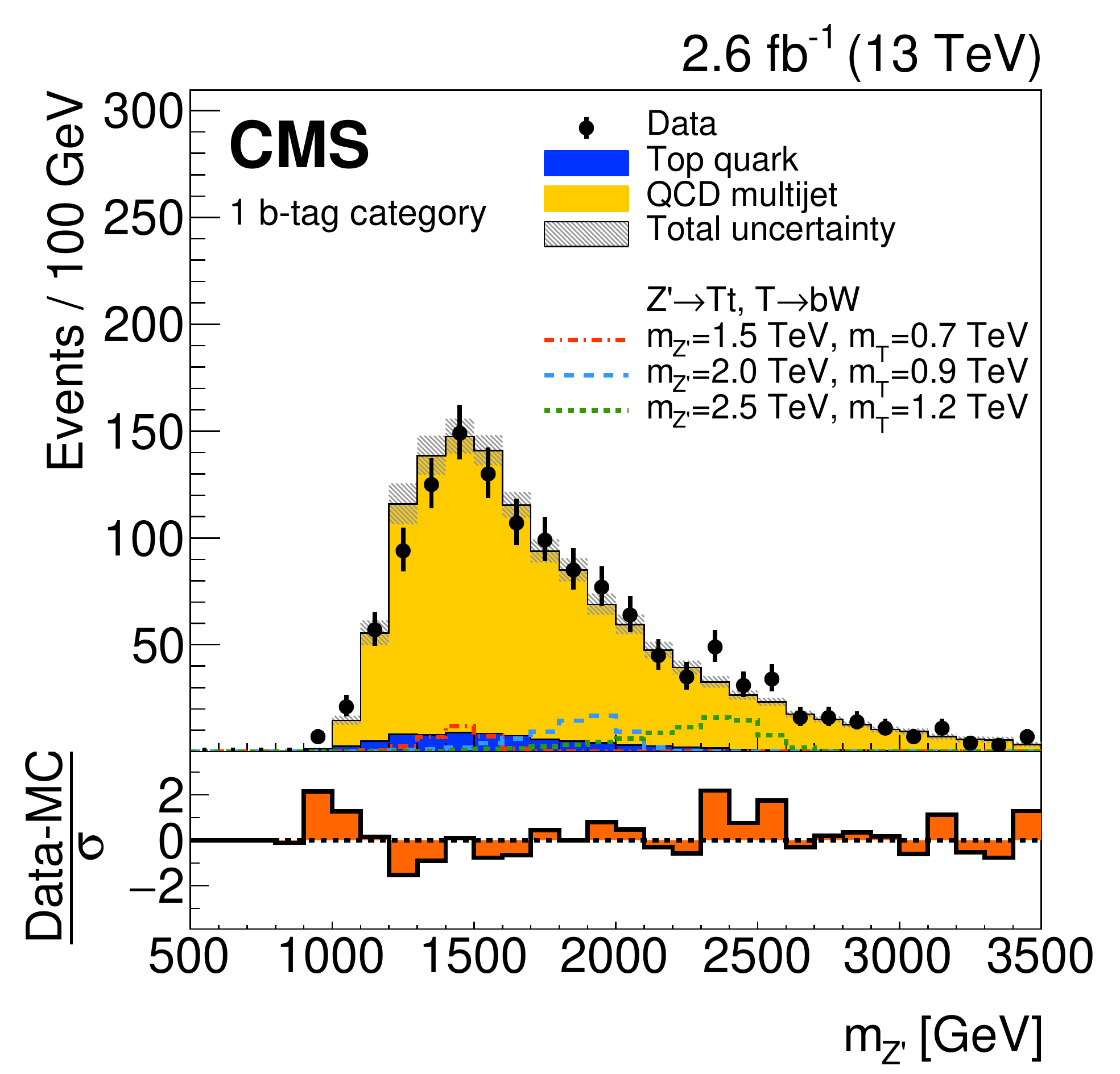}\\
\includegraphics[width=0.60\textwidth]{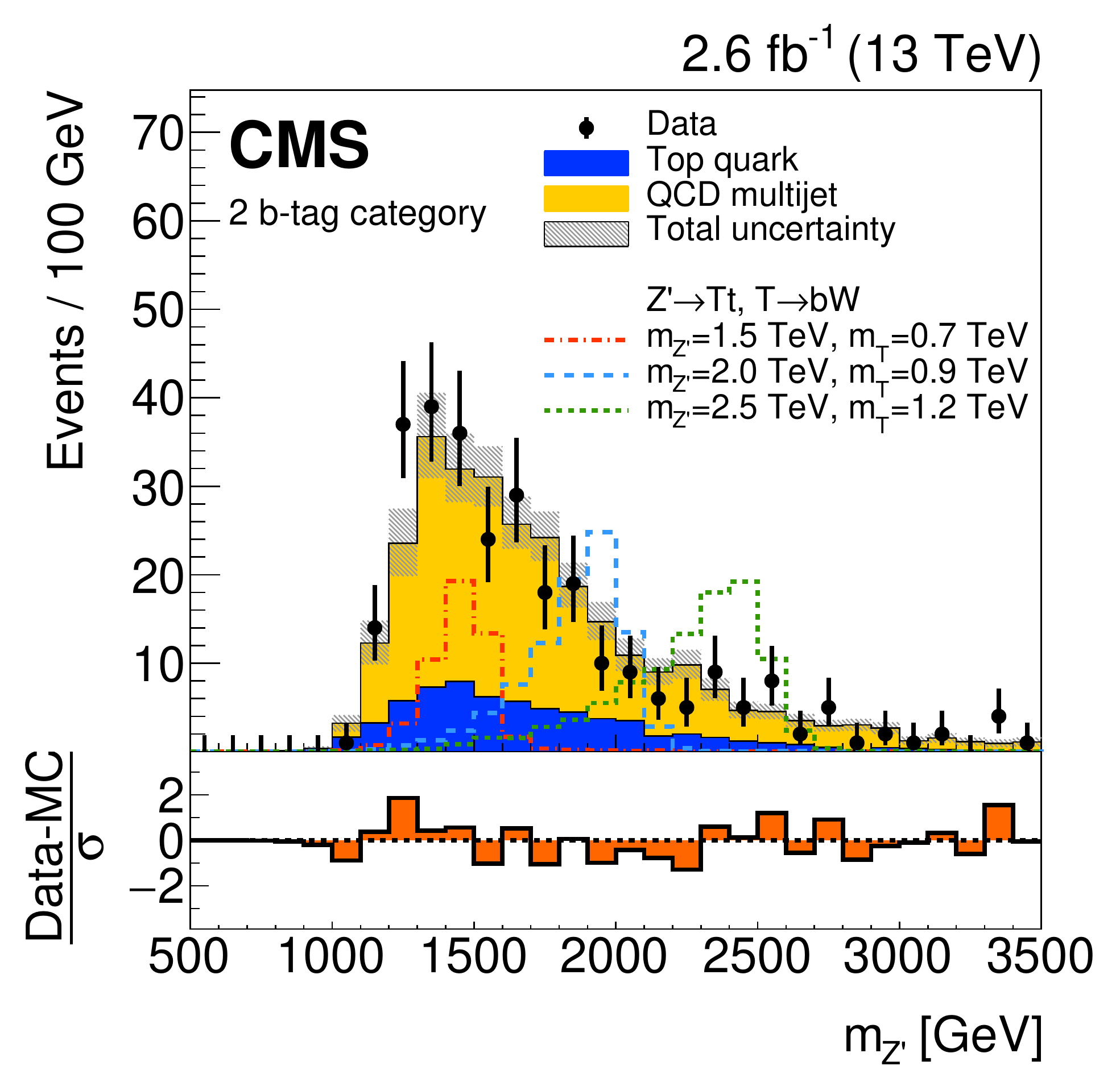}
\caption{Distribution of the $m_{\PZpr}$ variable for the signal region with 1 b tag (upper plot) and 2 b tags (lower plot) prior to the fit. The yellow (lighter) distribution represents the multijet background estimated from data, the blue (darker) distribution is the estimated top quark background, and the black markers are the data. The gray bands represent the statistical and systematic uncertainties in the background estimates. The uncertainty $\sigma$ includes the statistical uncertainties in data and backgrounds, and the systematic uncertainties in the estimated backgrounds. The dashed lines represent the distributions for signal hypotheses as indicated in the legend. The signal distributions are each normalized to a cross section of 1\unit{pb}. Events lying outside the x-axis range are not considered.}
\label{sr_final}
\end{figure}

\begin{figure}[htbp]
\centering
\includegraphics[width=0.49\textwidth]{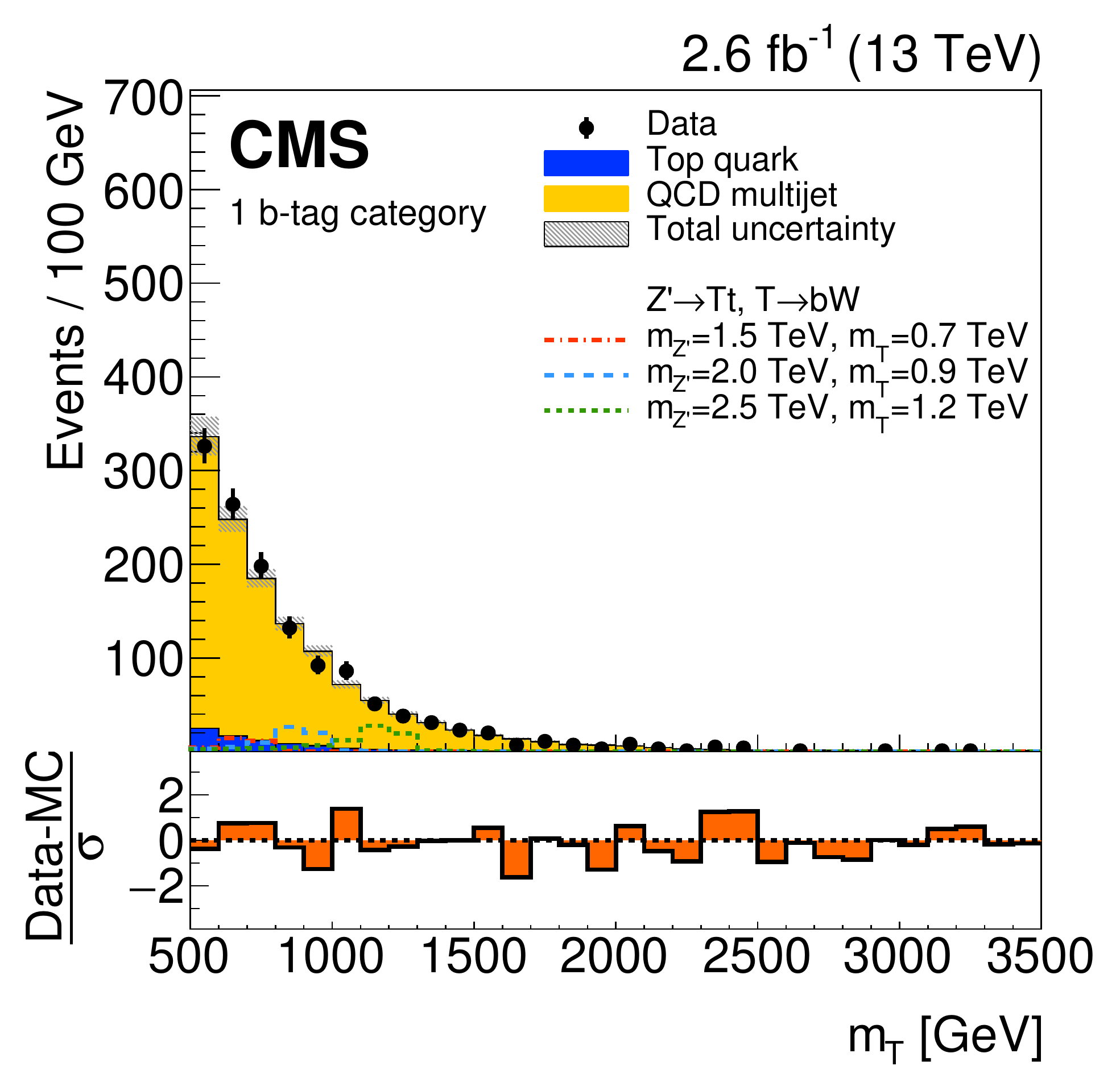}
\includegraphics[width=0.49\textwidth]{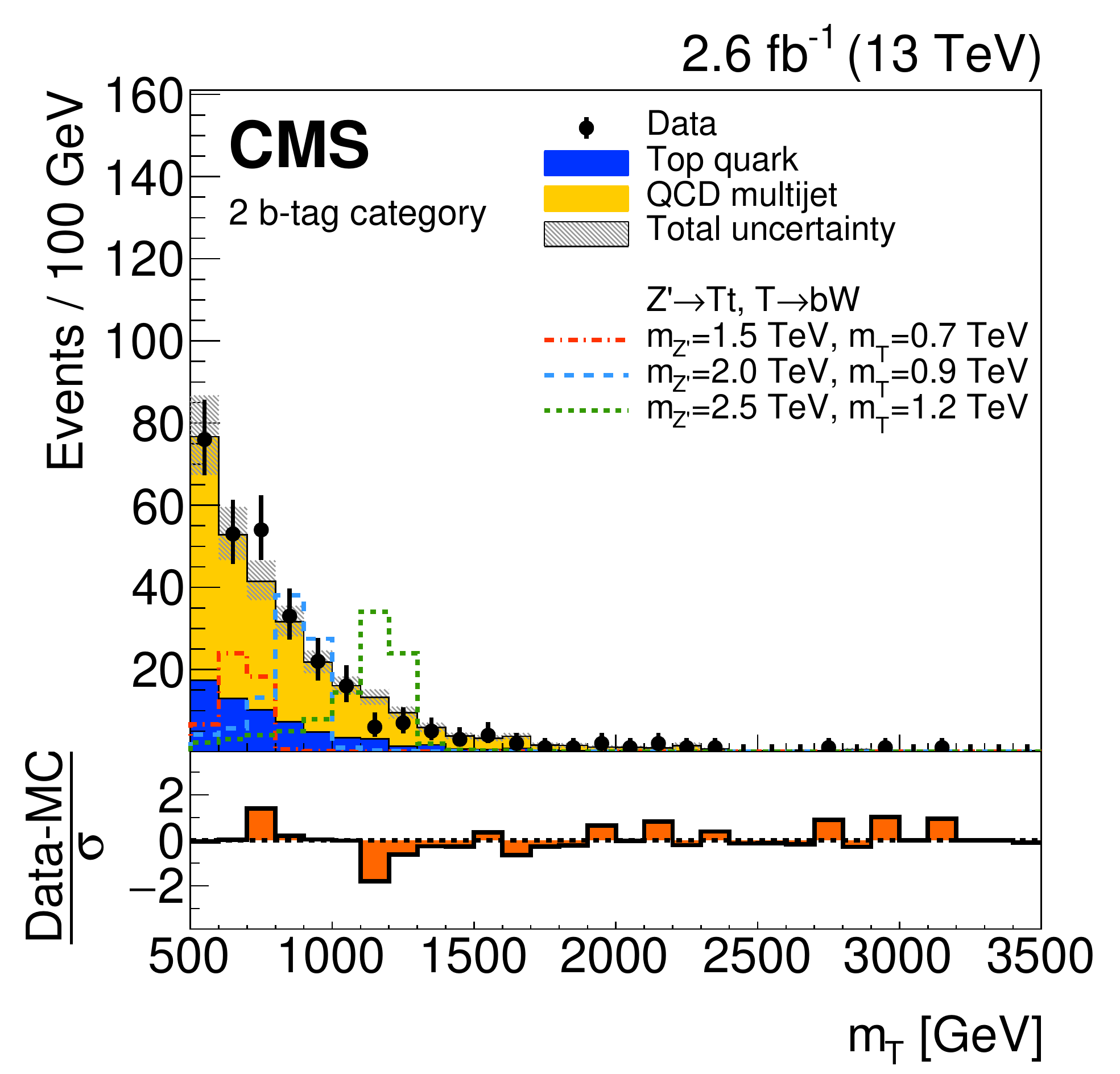}\\
\includegraphics[width=0.49\textwidth]{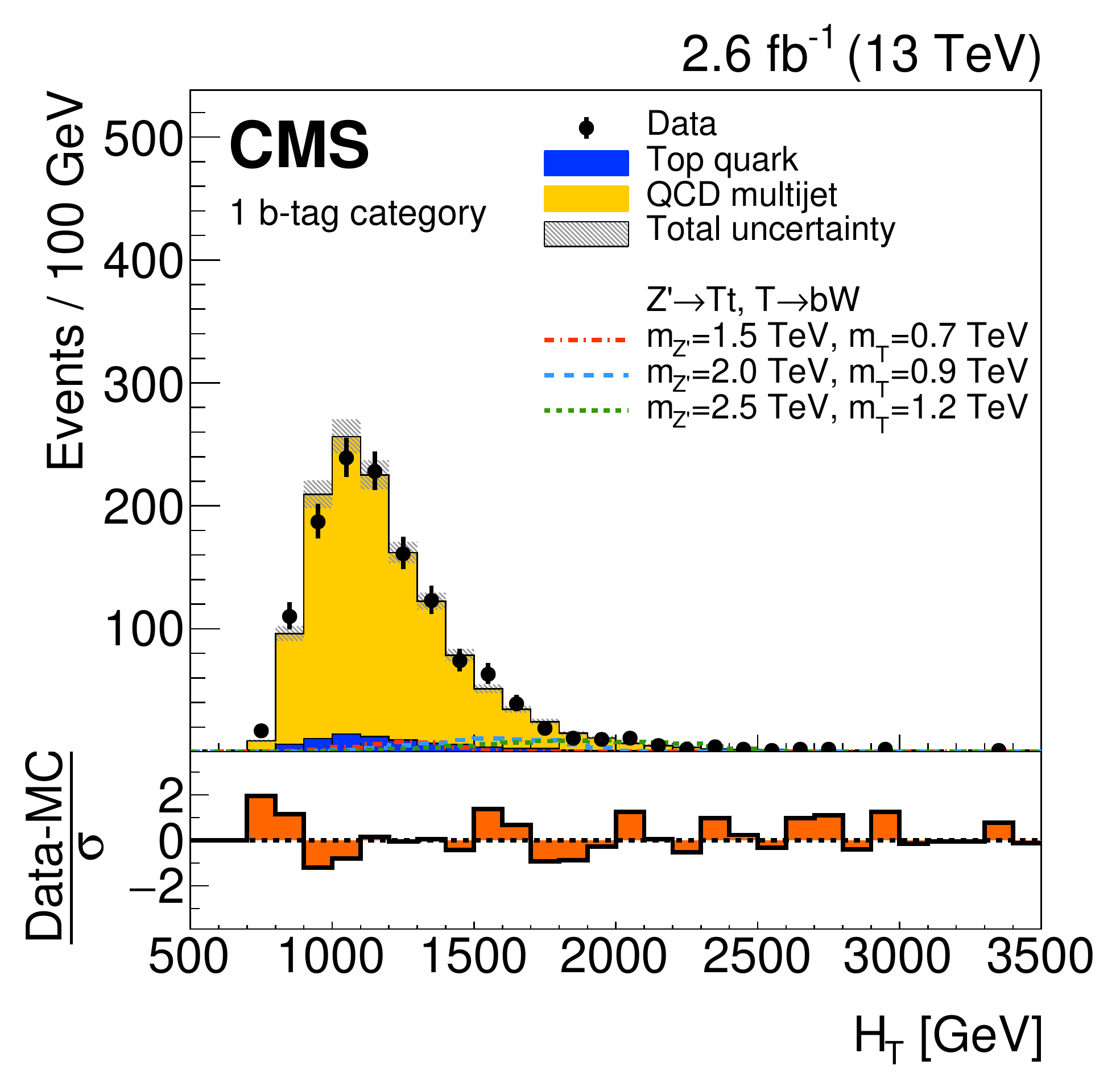}
\includegraphics[width=0.49\textwidth]{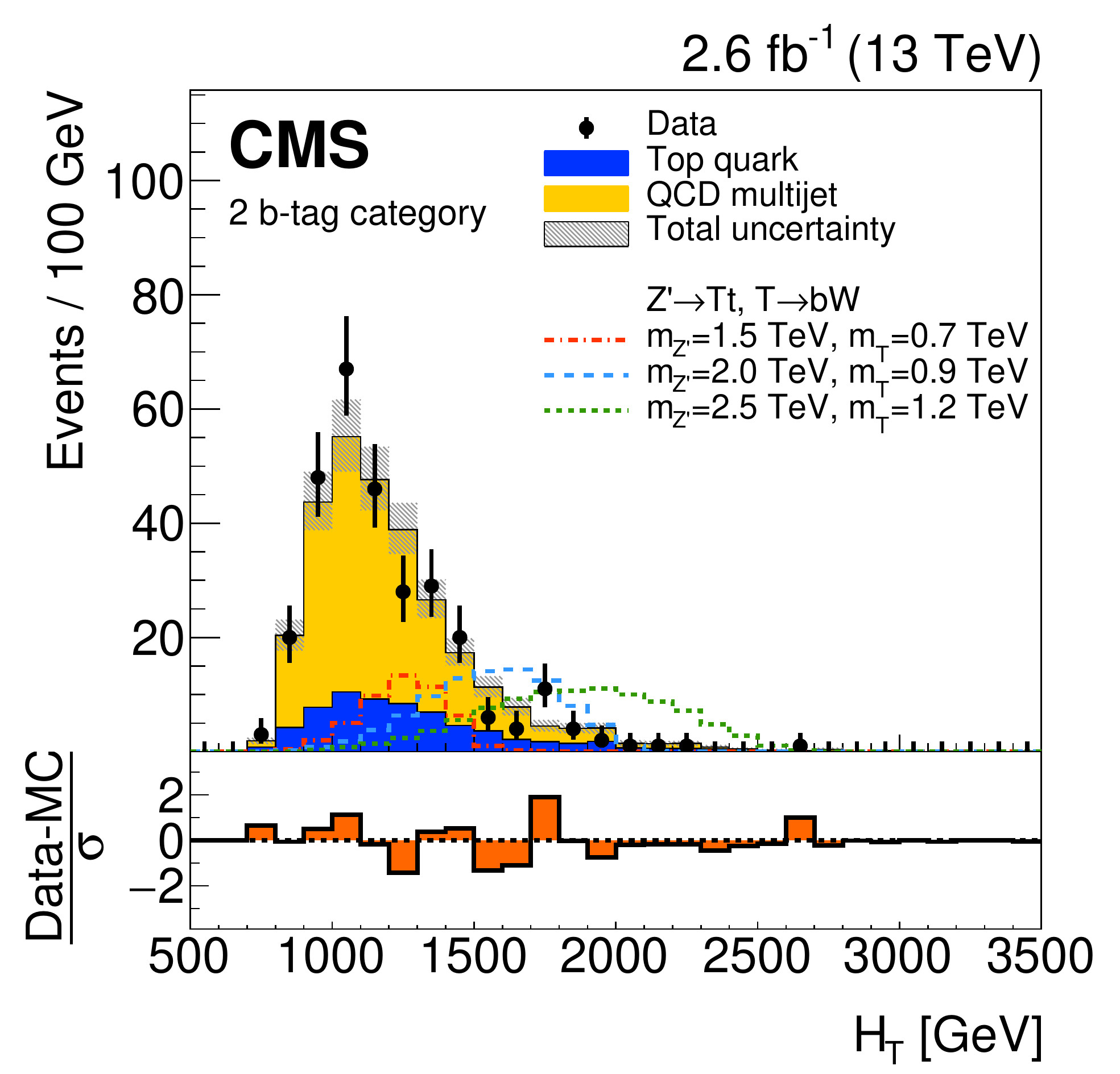}\\
\caption{Distributions of the $m_{\PQT}$ (upper plots) and $H_{\PQT}$ (lower plots) variables for the 1 b tag (left) and 2 b tag (right) event categories prior to the fit. The gray bands represent the statistical and systematic uncertainties in the background estimates. The uncertainty $\sigma$ includes the statistical uncertainties in data and backgrounds, and the systematic uncertainties in the estimated backgrounds. The dashed lines represent the distributions for signal hypotheses as indicated in the legend. The signal distributions are each normalized to a cross section of 1\unit{pb}.  Events lying outside the x-axis range are not considered.}
\label{sr_mtpht}
\end{figure}

The $m_{\PZpr}$ distributions in the two signal categories are shown in Fig.~\ref{sr_final}. The $m_{\PQT}$ and $H_{\PQT}$ variables are shown in Fig. \ref{sr_mtpht}. No excess with respect to the expected background is observed.

A template-based shape analysis with the  \textsc{Theta} software package \cite{thetaWeb} is performed, using the $m_{\PZpr}$ distribution in the two categories, to extract upper cross section limits on a hypothetical signal production. A Bayesian likelihood-based method \cite{Olive:2016xmw} is used. Expected limit intervals at 95\% confidence level (CL) are obtained by performing a large number of pseudo-experiments. The expected background model is varied within the systematic and statistical uncertainties to determine the best fit to the observed data. The modeling of uncertainties in the shapes  is performed through cubic-linear template morphing, where the cubic interpolation is used up to one standard deviation and the linear interpolation beyond that \cite{thetaWeb}.  A nuisance parameter is assigned for each systematic uncertainty in the likelihood. For the parameter of interest, i.e. the signal cross section, a uniform prior is used, while log-normal priors are used for the nuisance parameters. The two event categories are fitted simultaneously.

To avoid the normalization of the multijet background being affected by the presence of a hypothetical signal, a prior uncertainty of 50\% is assigned in the fit of the signal hypothesis, as discussed in Section \ref{background}. The fit is able to constrain the multijet normalization, primarily with the 1 b tag category, which has more events and is less signal enriched.

Table~\ref{limittableWB} shows the expected and observed limits on the cross section to produce a $\PZpr$ boson that decays to Tt for different $\PZpr$ boson and T quark mass hypotheses. Three different hypotheses for the decay of the T quark are considered: 100\% branching fraction into bW, tH, or tZ.
The effect of increasing the width of the $\PZpr$ boson or the T quark to 10\% on a single signal configuration has been studied and the impact on the cross section limits is found to be negligible in both cases, because of the detector resolution being bigger than 15\%.

\begin{table}
\centering
\topcaption{Table of expected and observed limits on the cross section to produce a $\PZpr$ boson that decays to Tt at 95\% CL for the $\mathrm{T}\to\mathrm{bW}$ (upper), $\mathrm{T}\to\mathrm{tH}$ (middle), and $\mathrm{T}\to\mathrm{tZ}$ (lower) signal hypotheses.}
\label{limittableWB}
\begin{tabular}{cccccccc}
\hline
\multicolumn{8}{c}{${\mathcal{B}(\PQT\to \PQb\PW)}=1$}\\
\hline
$m_{\PZpr}$ [\GeVns{}] & $m_{\PQT}$ [\GeVns{}] & Observed [pb] & \multicolumn{5}{c}{Expected [pb]} \\\cline{4-8}
 & & & $-2\sigma$ & $-1\sigma$ & Median & $+1\sigma$ & $+2\sigma$ \\
\hline
1500 & 700 & 0.73 & 0.32 & 0.48 & 0.67 & 1.0 & 1.6 \\
1500 & 900 & 1.5 & 0.64 & 0.94 & 1.5 & 2.2 & 3.7 \\
1500 & 1200 & 8.6 & 3.7 & 5.2 & 7.8 & 13 & 22 \\
2000 & 900 & 0.19 & 0.17 & 0.24 & 0.36 & 0.56 & 0.90 \\
2000 & 1200 & 0.27 & 0.24 & 0.33 & 0.49 & 0.76 & 1.3 \\
2000 & 1500 & 0.96 & 0.82 & 1.2 & 1.9 & 3.0 & 5.4 \\
2500 & 1200 & 0.29 & 0.10 & 0.15 & 0.24 & 0.39 & 0.64 \\
2500 & 1500 & 0.30 & 0.11 & 0.16 & 0.24 & 0.39 & 0.65 \\[2ex]
\hline
\multicolumn{8}{c}{${\mathcal{B}(\PQT\to \PQt\PH)}=1$}\\
\hline
$m_{\PZpr}$ [\GeVns{}] & $m_{\PQT}$ [\GeVns{}] & Observed [pb] & \multicolumn{5}{c}{Expected [pb]} \\\cline{4-8}
 & & & $-2\sigma$ & $-1\sigma$ & Median & $+1\sigma$ & $+2\sigma$ \\
\hline
1500 & 700 & 4.0 & 0.98 & 1.4 & 2.1 & 3.3 & 5.8 \\
1500 & 900 & 3.2 & 0.76 & 1.0 & 1.6 & 2.6 & 4.2 \\
1500 & 1200 & 9.4 & 2.6 & 3.6 & 5.6 & 9.3 & 19 \\
2000 & 900 & 0.53 & 0.39 & 0.55 & 0.84 & 1.4 & 2.3 \\
2000 & 1200 & 0.53 & 0.36 & 0.52 & 0.79 & 1.2 & 2.2 \\
2000 & 1500 & 0.60 & 0.50 & 0.67 & 0.99 & 1.6 & 2.9 \\
2500 & 1200 & 0.24 & 0.24 & 0.34 & 0.52 & 0.83 & 1.5 \\
2500 & 1500 & 0.23 & 0.21 & 0.31 & 0.49 & 0.81 & 1.3 \\[2ex]
\hline
\multicolumn{8}{c}{${\mathcal{B}(\PQT\to \PQt\Z)}=1$}\\
\hline
$m_{\PZpr}$ [\GeVns{}] & $m_{\PQT}$ [\GeVns{}] & Observed [pb] & \multicolumn{5}{c}{Expected [pb]}\\\cline{4-8}
 & & & $-2\sigma$ & $-1\sigma$ & Median & $+1\sigma$ & $+2\sigma$ \\
\hline
1500 & 700 & 3.1 & 0.84 & 1.2 & 1.8 & 2.9 & 4.7 \\
1500 & 900 & 2.8 & 0.77 & 1.1 & 1.6 & 2.5 & 4.3 \\
1500 & 1200 & 3.4 & 1.3 & 1.8 & 2.7 & 4.2 & 6.4 \\
2000 & 900 & 0.37 & 0.30 & 0.41 & 0.61 & 0.97 & 1.8 \\
2000 & 1200 & 0.30 & 0.23 & 0.34 & 0.50 & 0.80 & 1.3 \\
2000 & 1500 & 0.32 & 0.26 & 0.37 & 0.55 & 0.85 & 1.7 \\
2500 & 1200 & 0.16 & 0.14 & 0.21 & 0.31 & 0.52 & 0.92 \\
2500 & 1500 & 0.13 & 0.12 & 0.17 & 0.27 & 0.45 & 0.77 \\
\hline
\end{tabular}
\end{table}

\begin{figure}[htbp]
\centering
\includegraphics[width=0.42\textwidth]{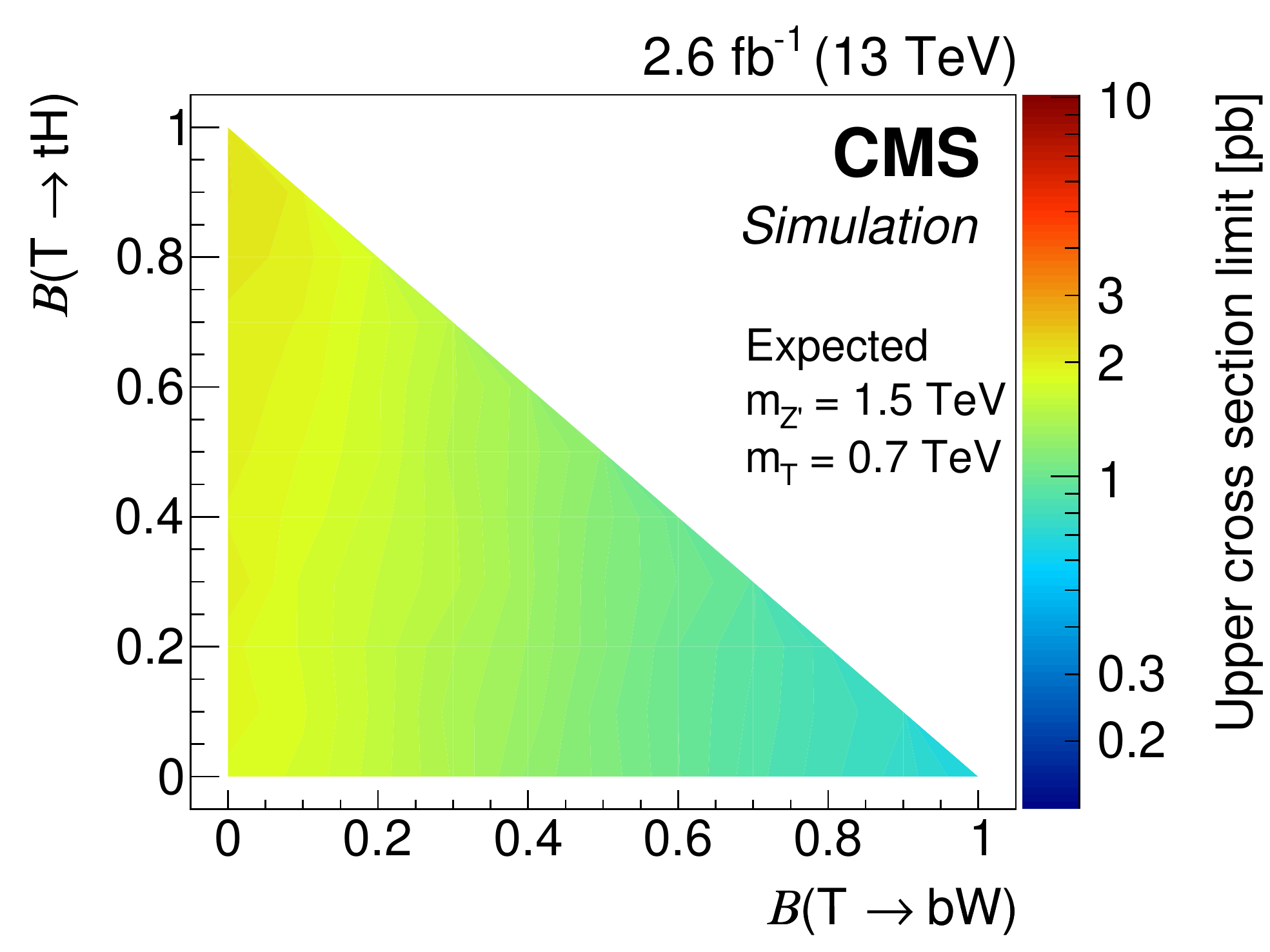}
\includegraphics[width=0.42\textwidth]{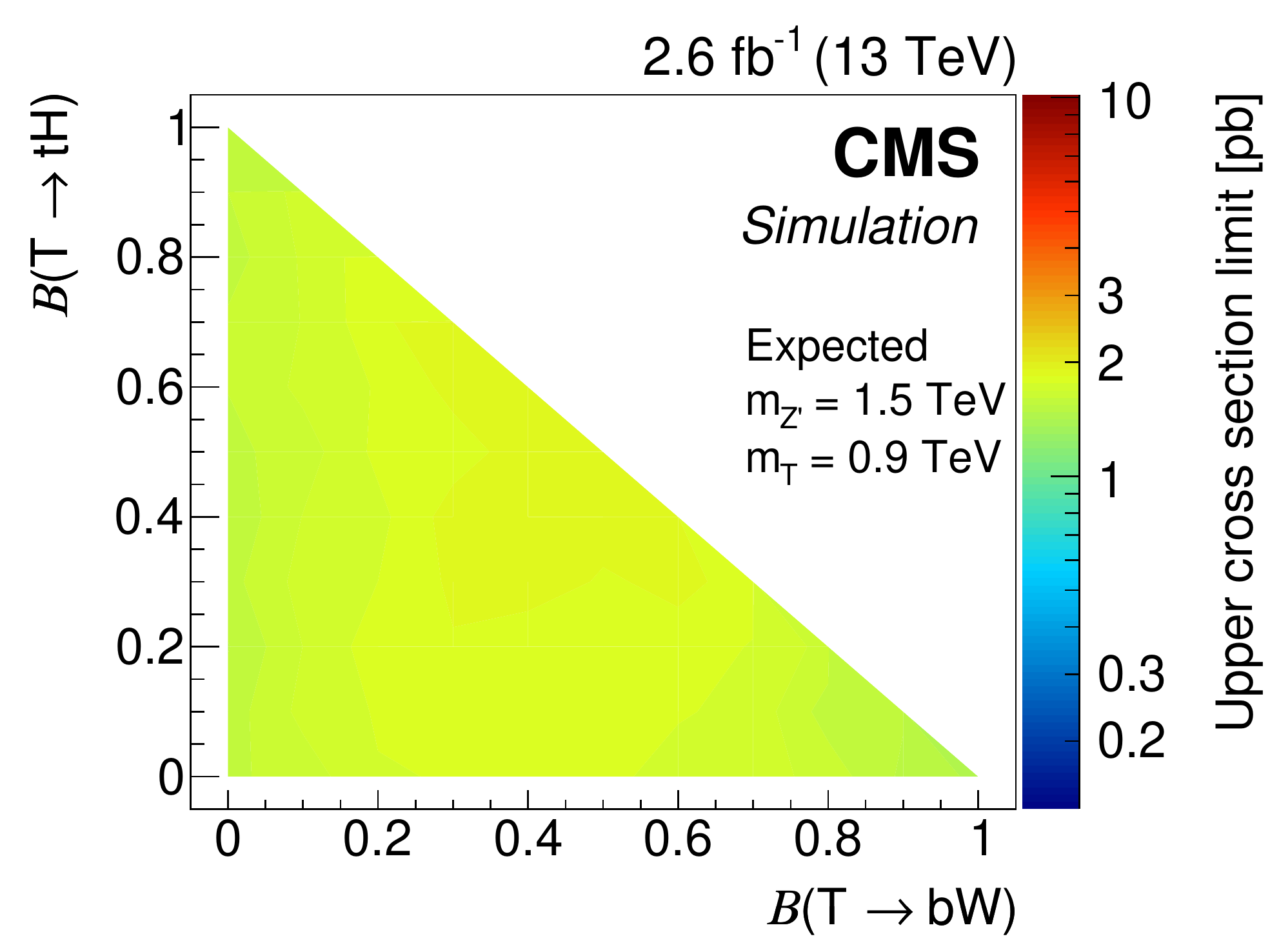}\\
\includegraphics[width=0.42\textwidth]{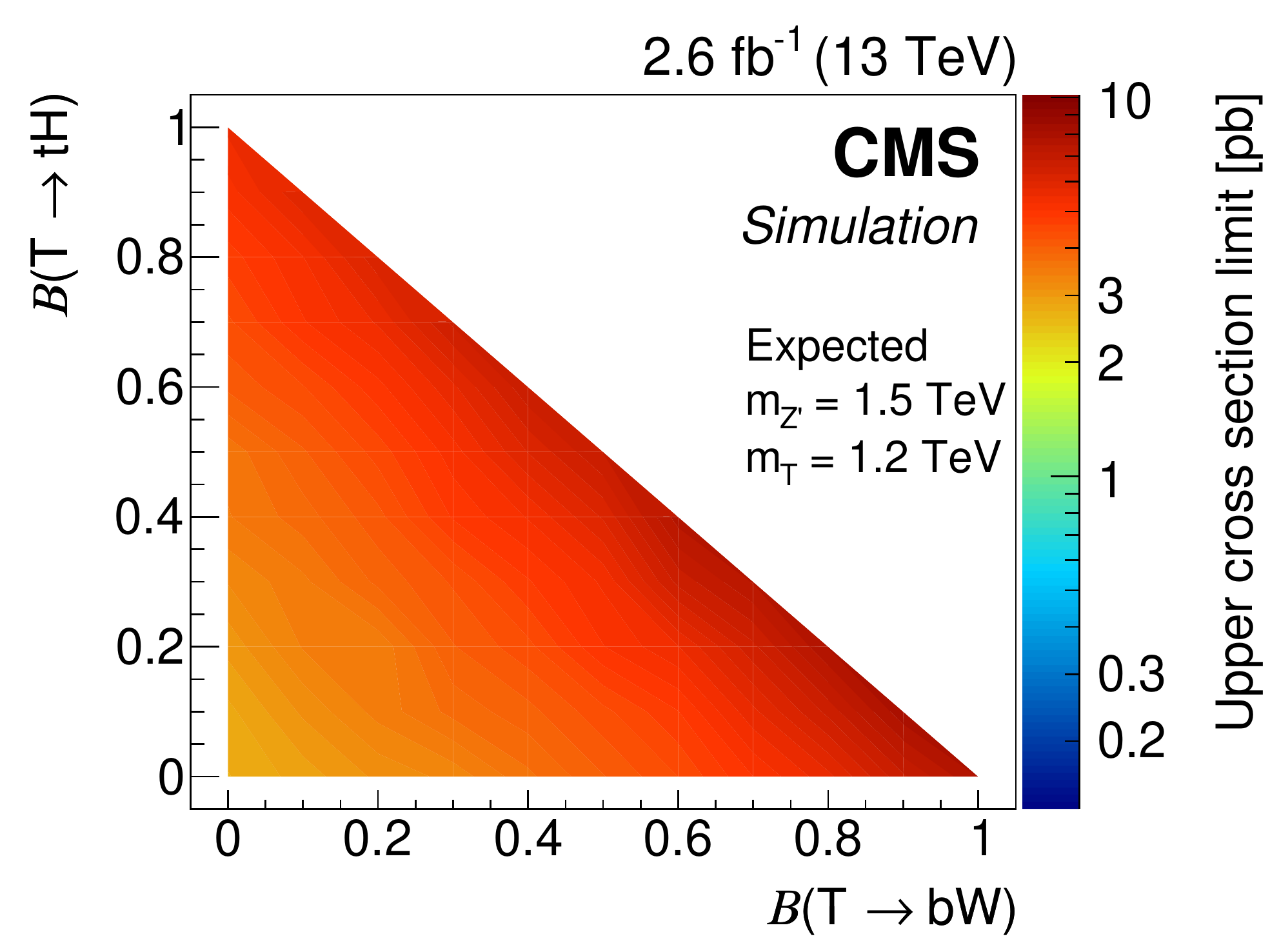}
\includegraphics[width=0.42\textwidth]{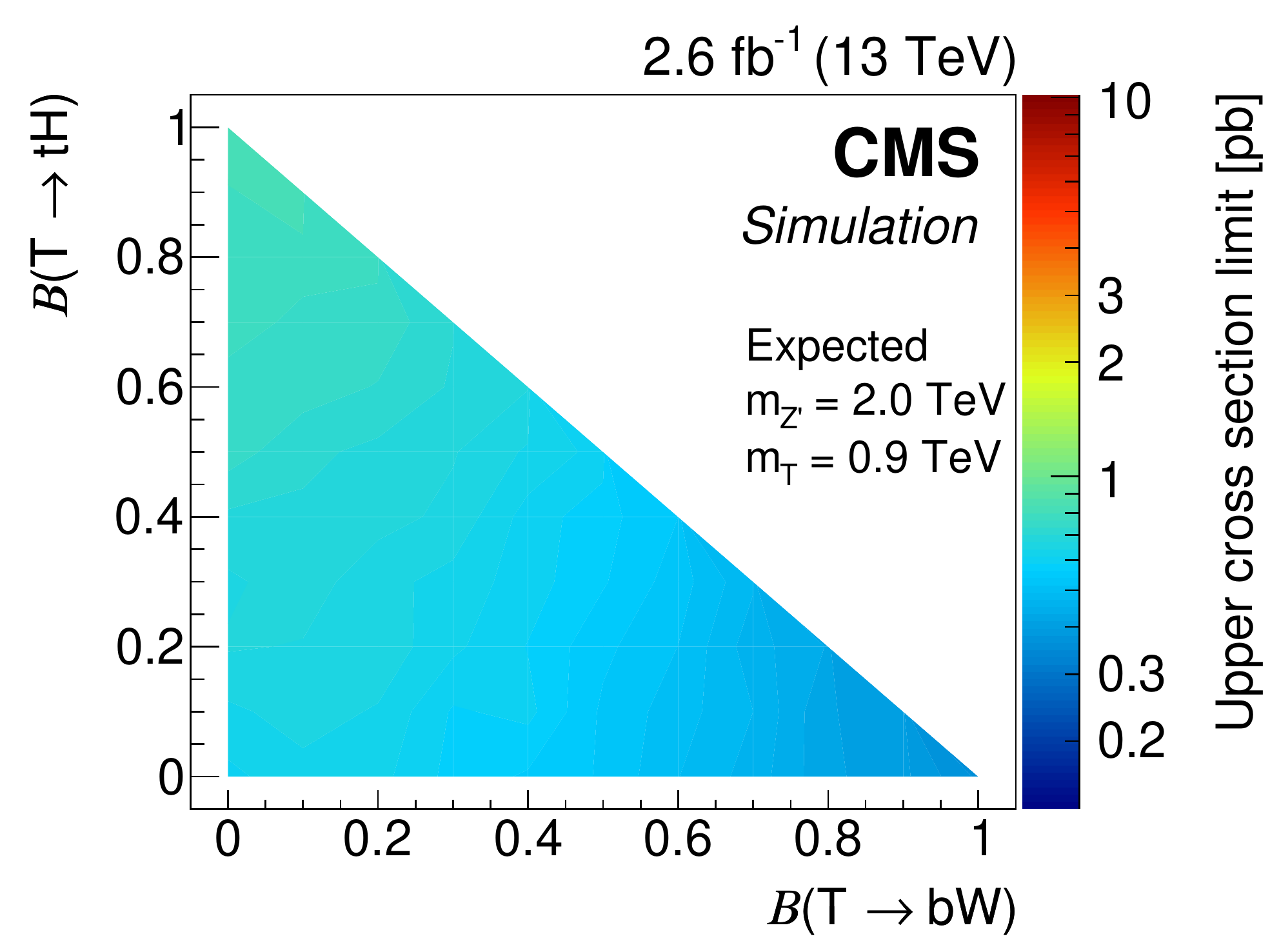}\\
\includegraphics[width=0.42\textwidth]{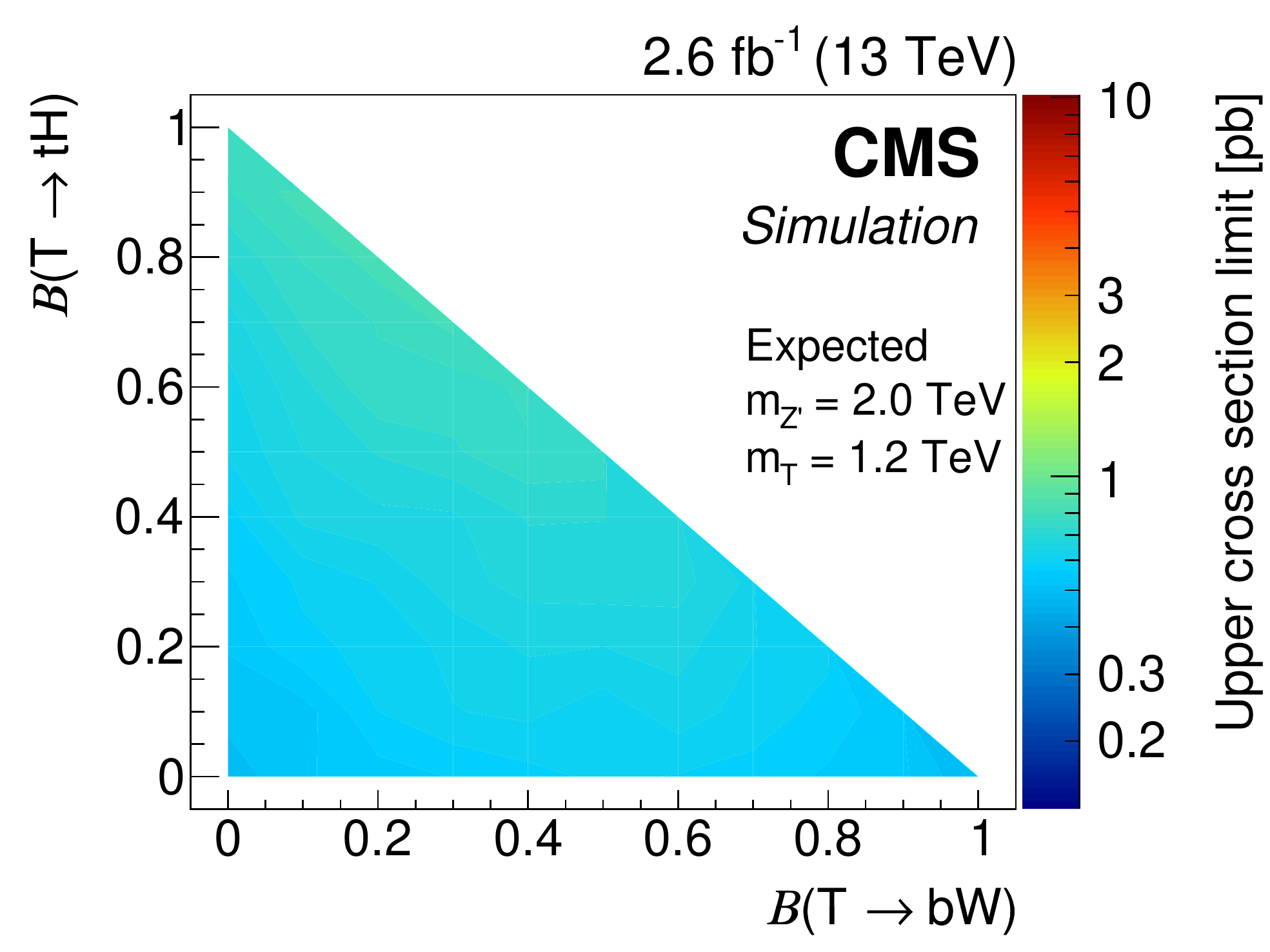}
\includegraphics[width=0.42\textwidth]{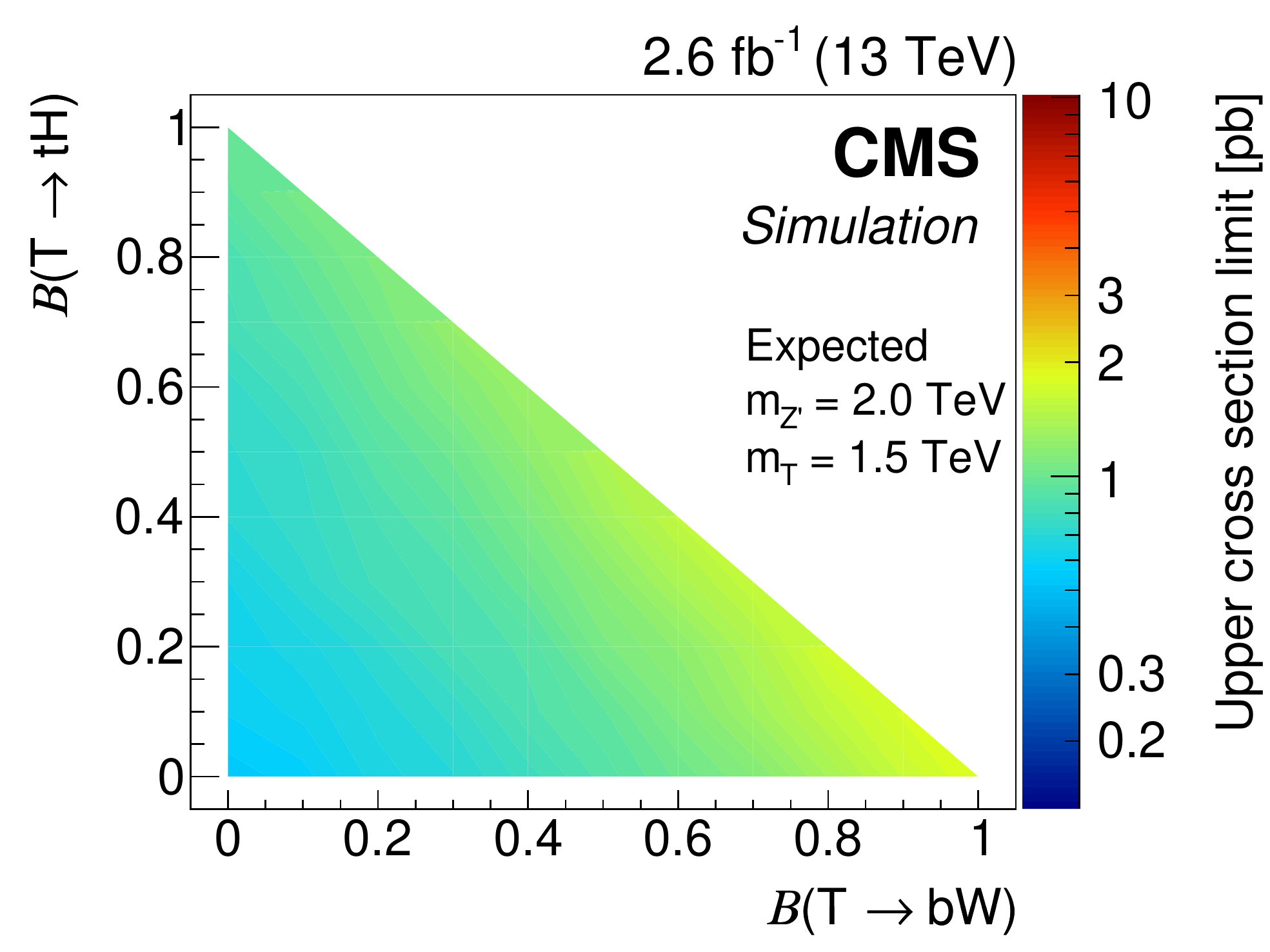}\\
\includegraphics[width=0.42\textwidth]{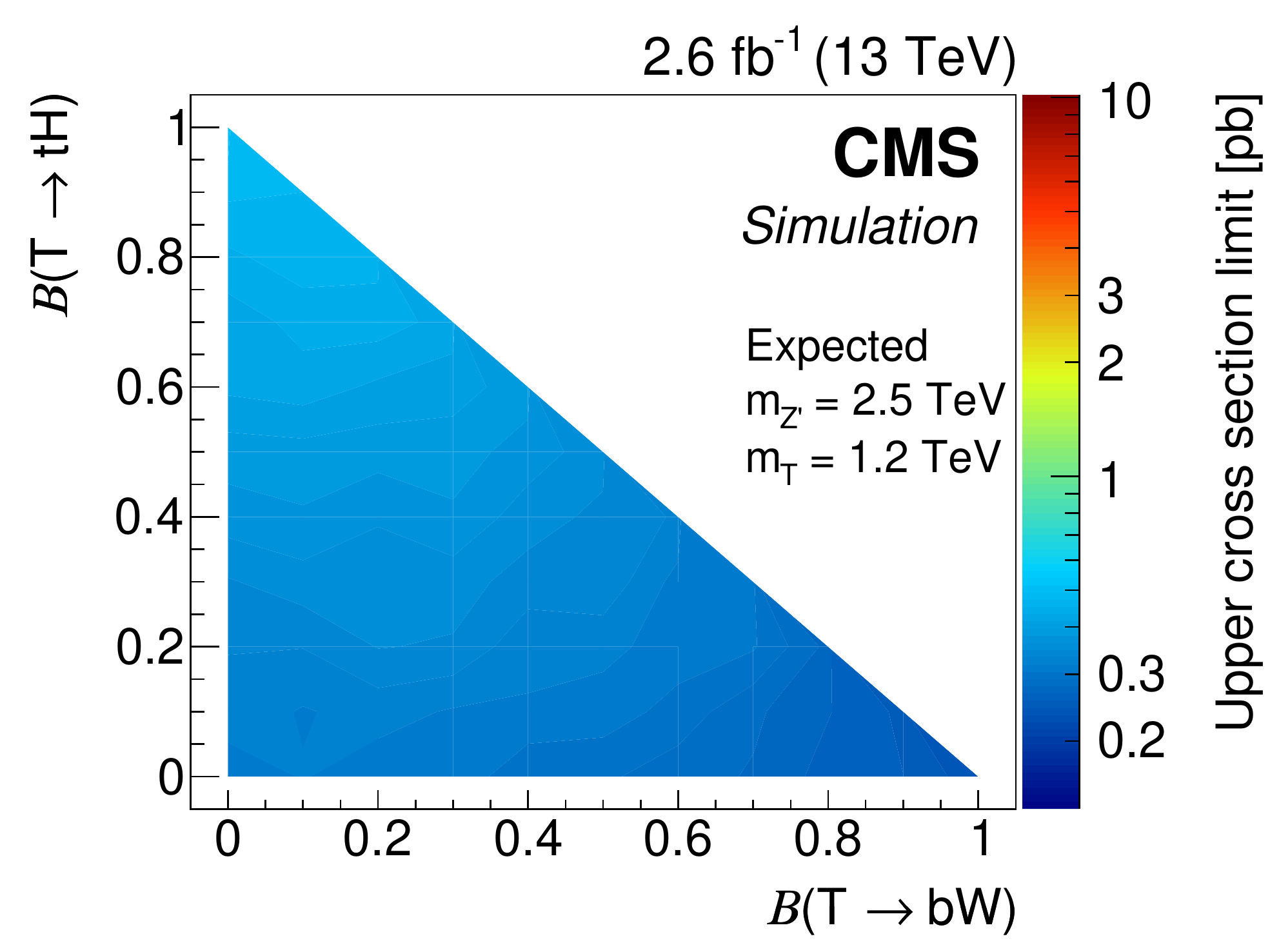}
\includegraphics[width=0.42\textwidth]{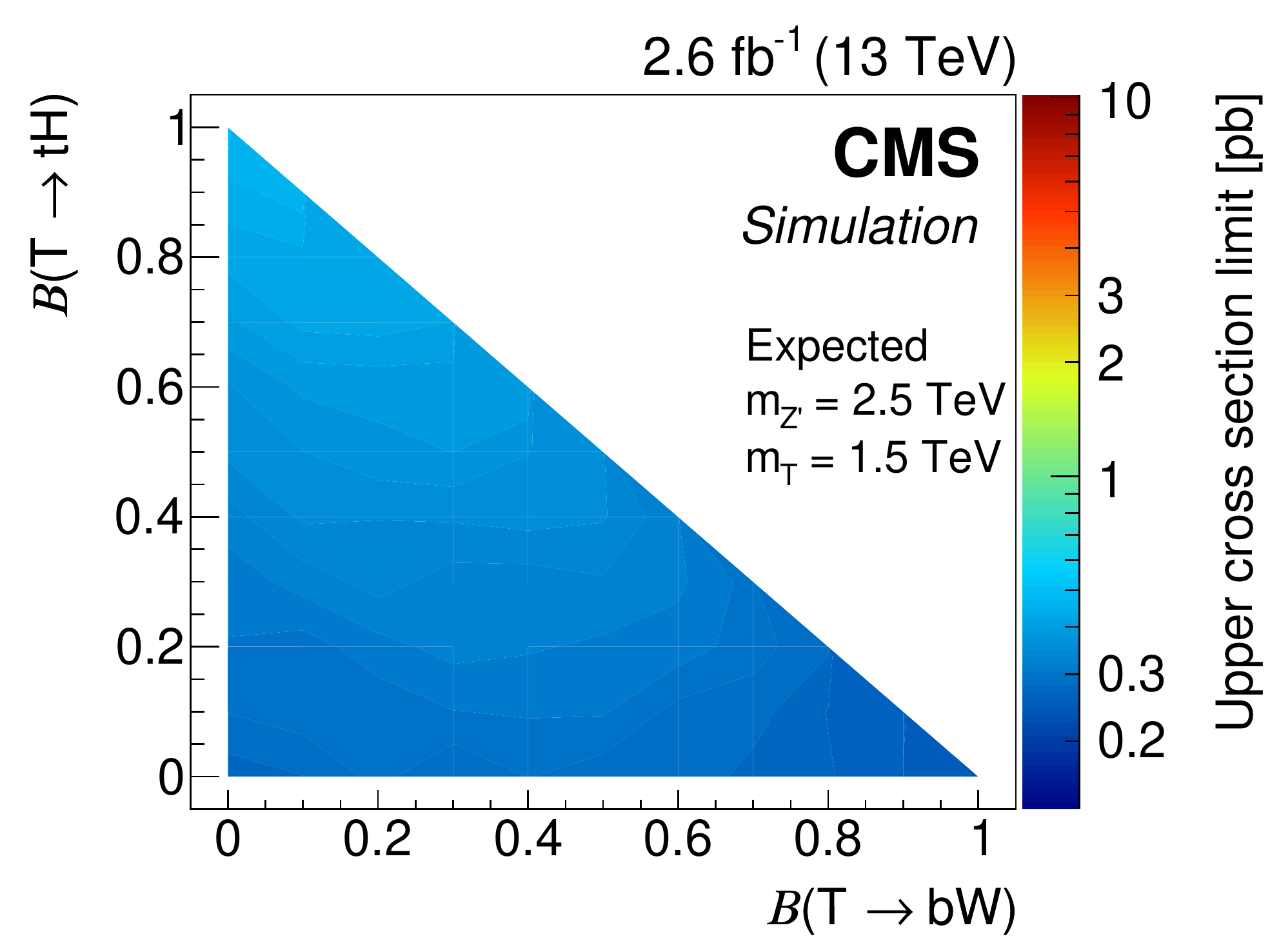}
\caption{Expected cross section limits for $\PZpr\to \PQT\PQt$ for different hypotheses for the $\PZpr$ boson and T quark masses, and the branching fraction of the T quark decay into bW and tH channels, with $\mathcal{B}(\PQT\to\PQt\Z)=(1-\mathcal{B}(\PQT\to \PQb\PW, \PQt\PH))$.
}
\label{exptriangle}
\end{figure}

\begin{figure}[htbp]
\centering
\includegraphics[width=0.42\textwidth]{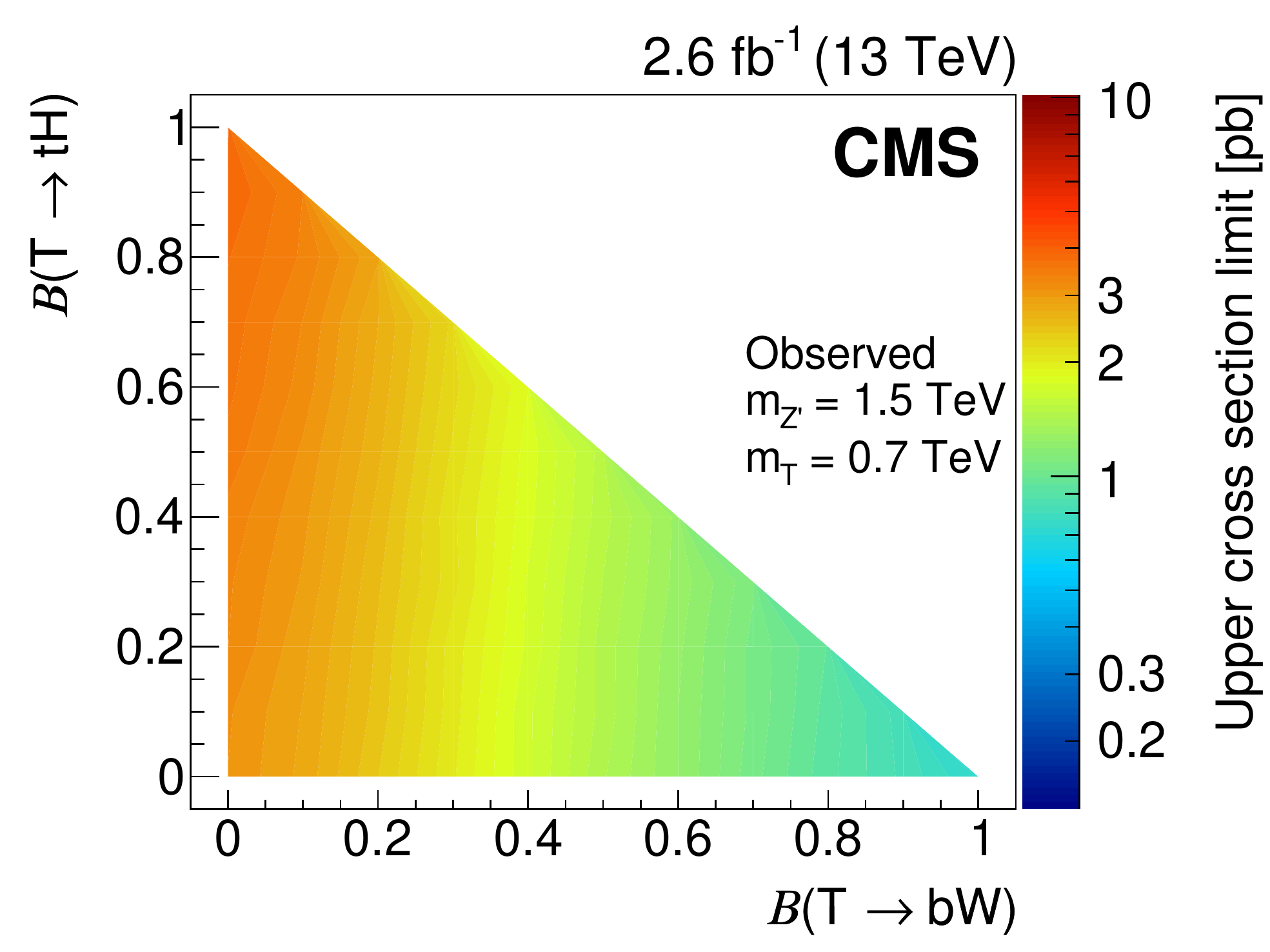}
\includegraphics[width=0.42\textwidth]{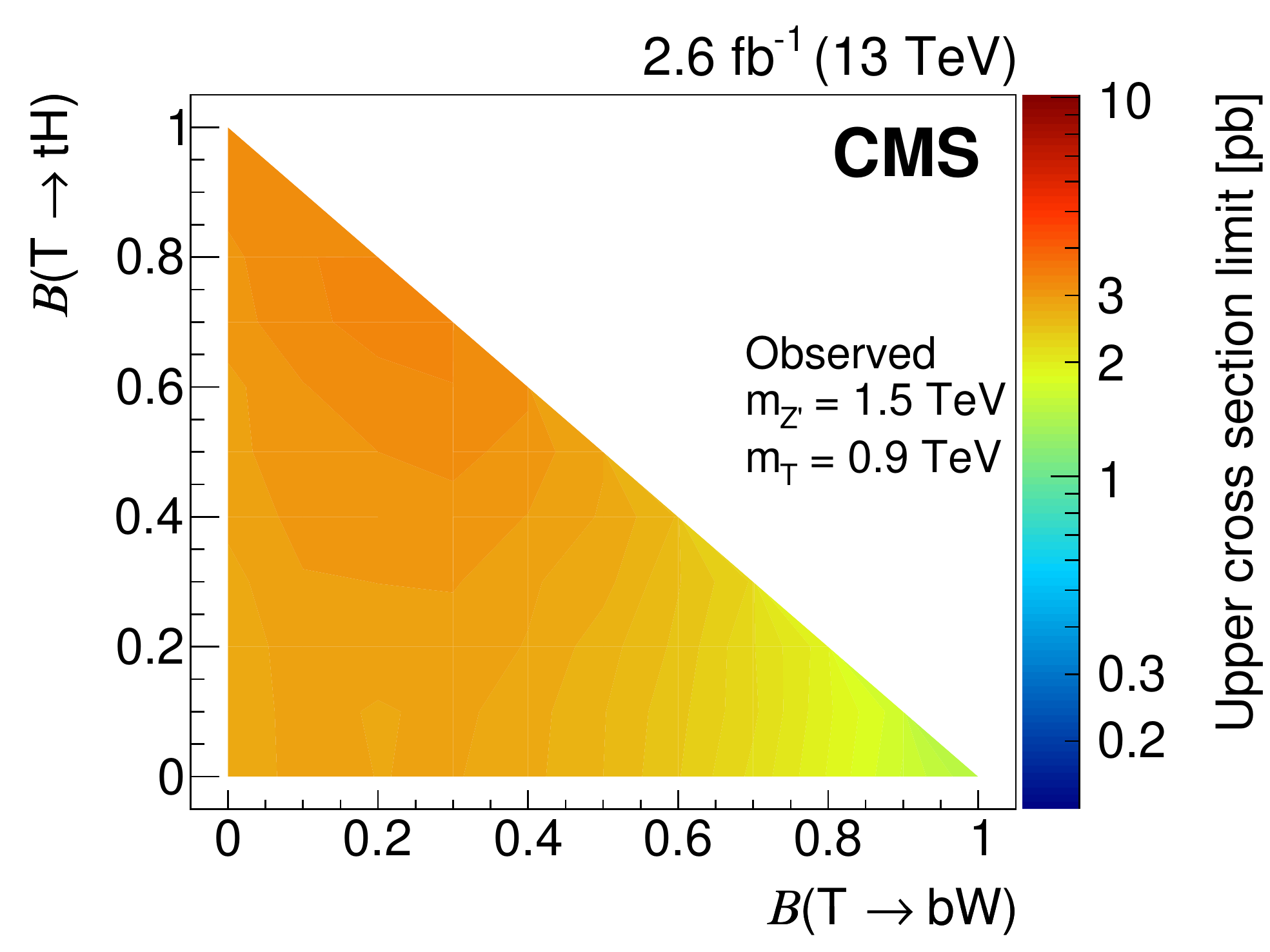}\\
\includegraphics[width=0.42\textwidth]{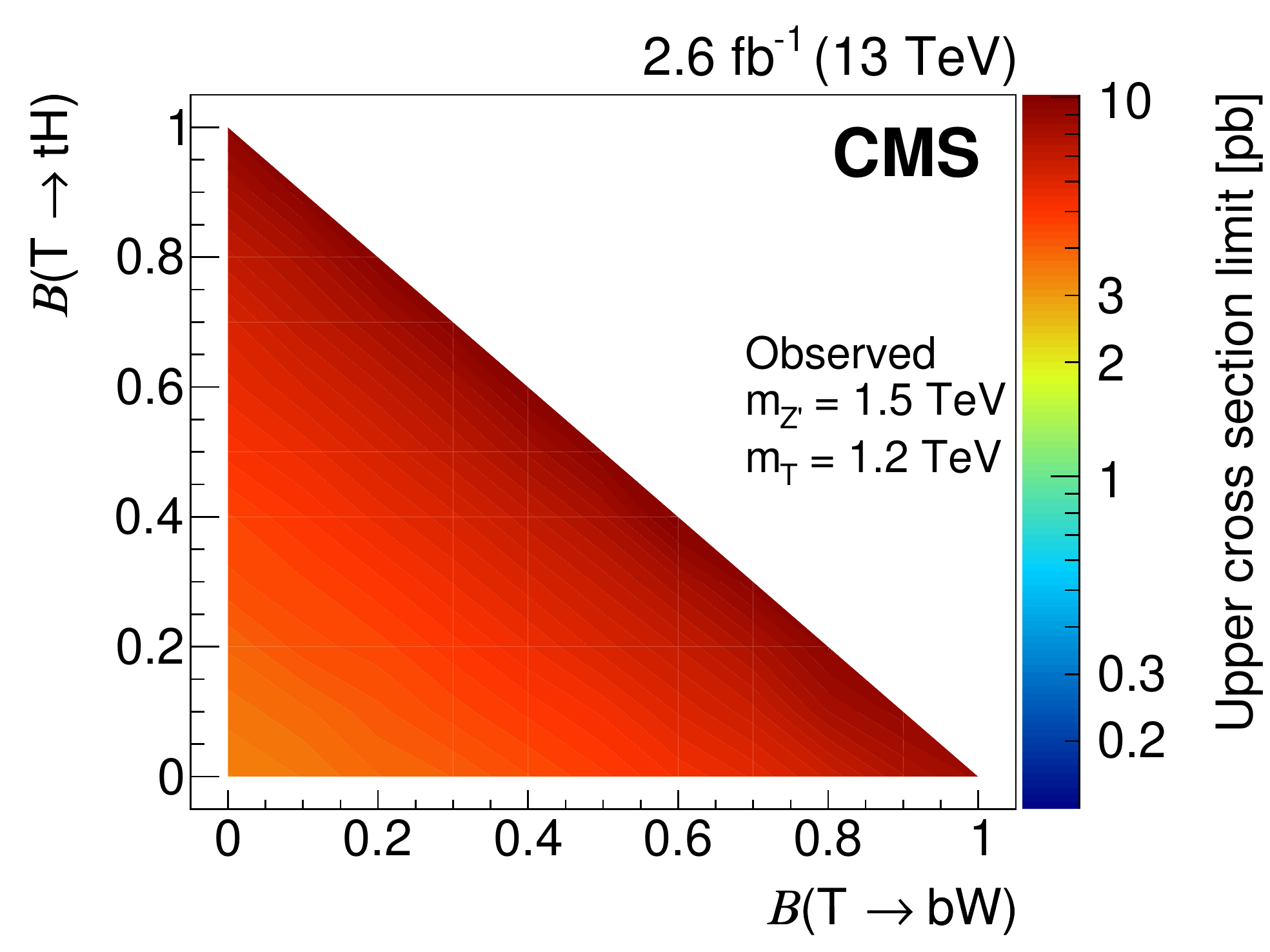}
\includegraphics[width=0.42\textwidth]{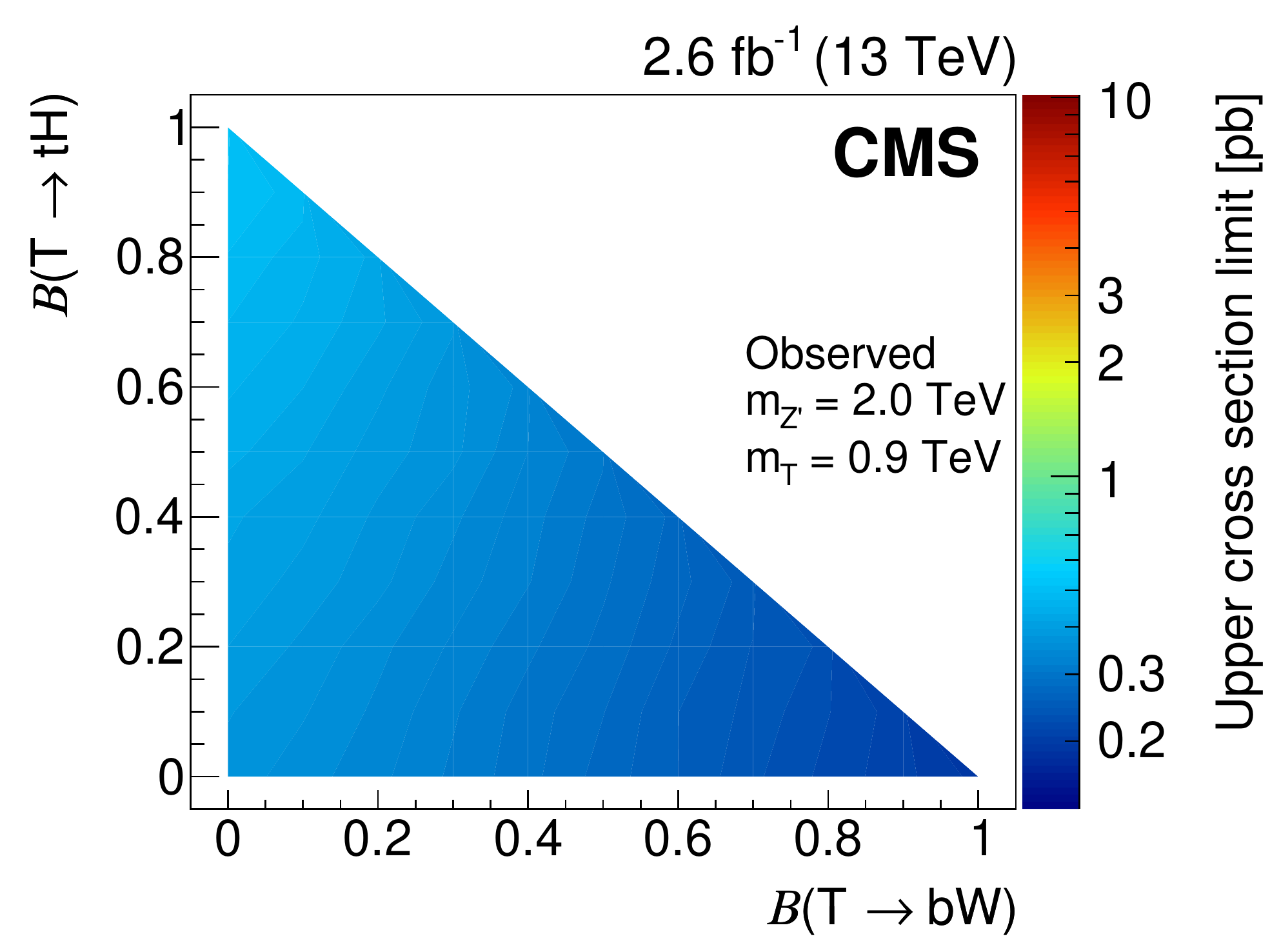}\\
\includegraphics[width=0.42\textwidth]{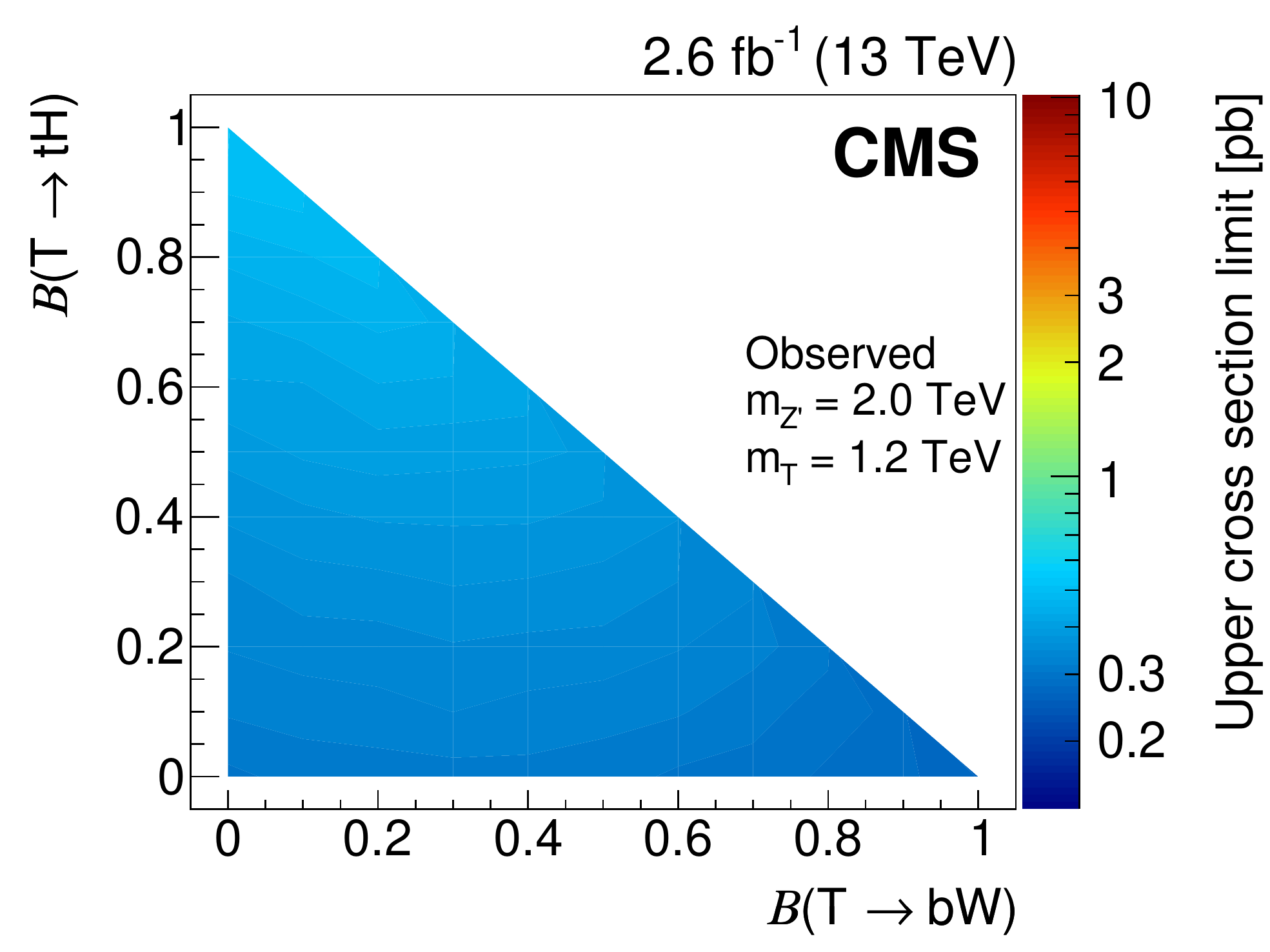}
\includegraphics[width=0.42\textwidth]{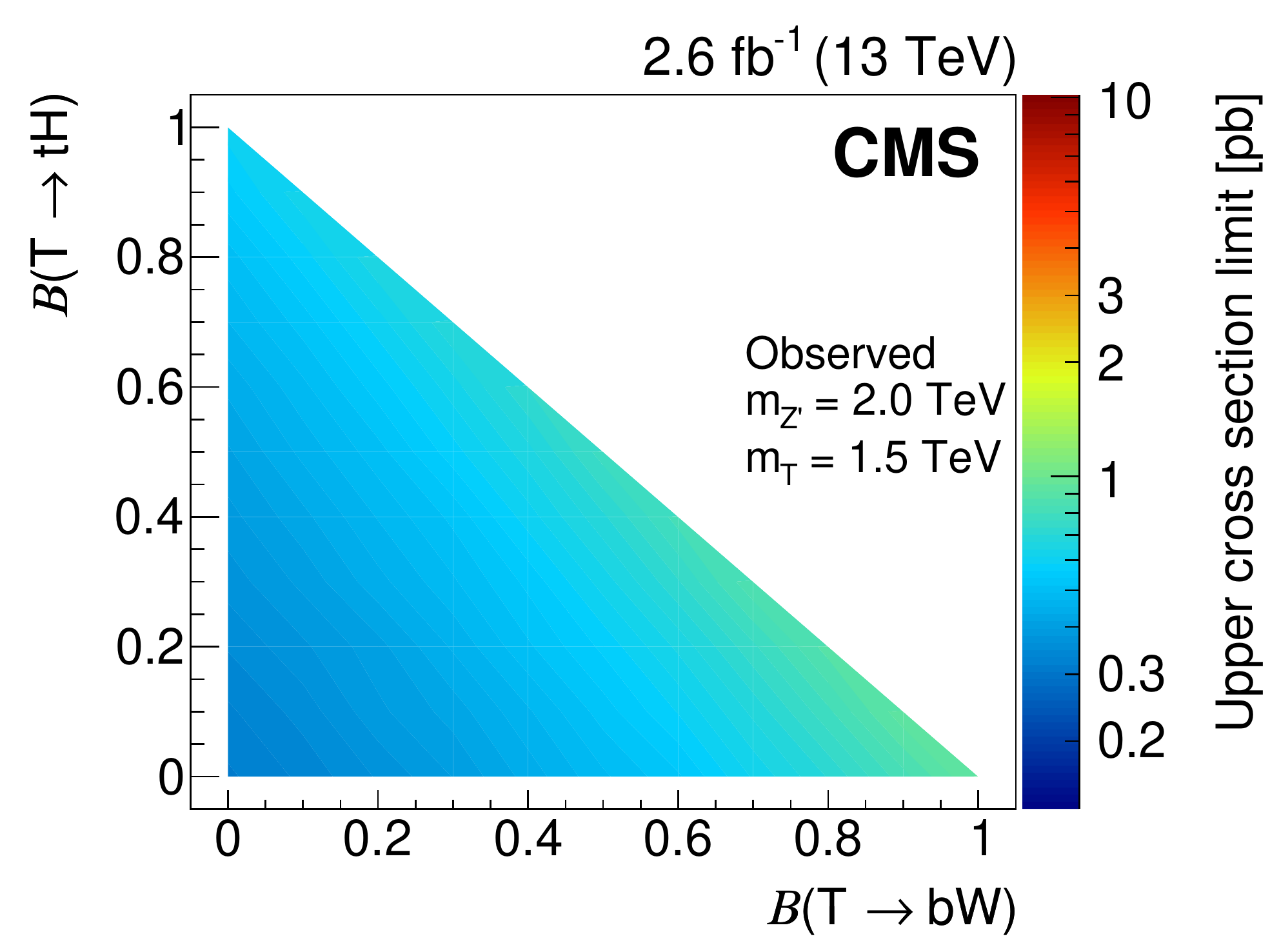}\\
\includegraphics[width=0.42\textwidth]{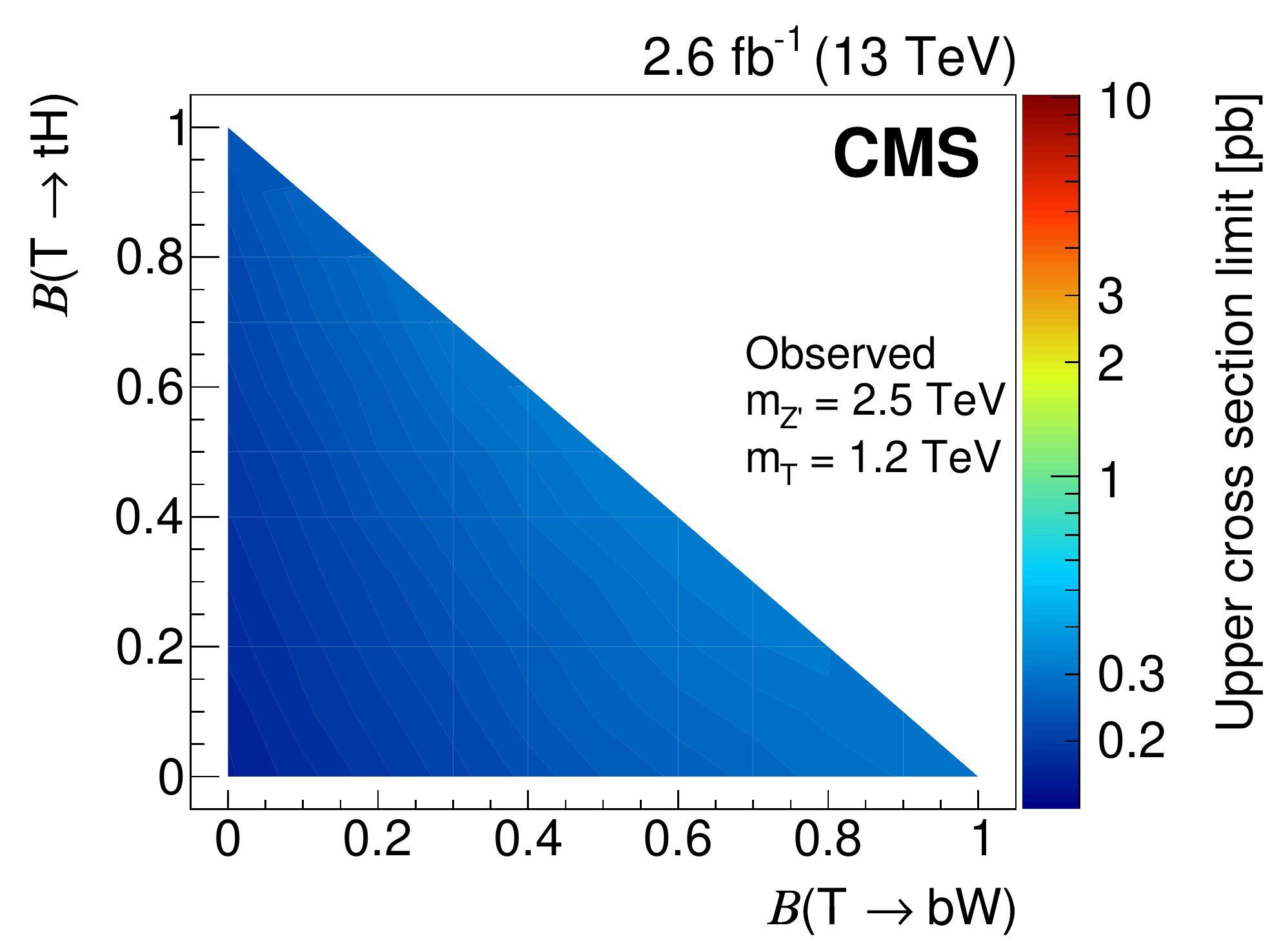}
\includegraphics[width=0.42\textwidth]{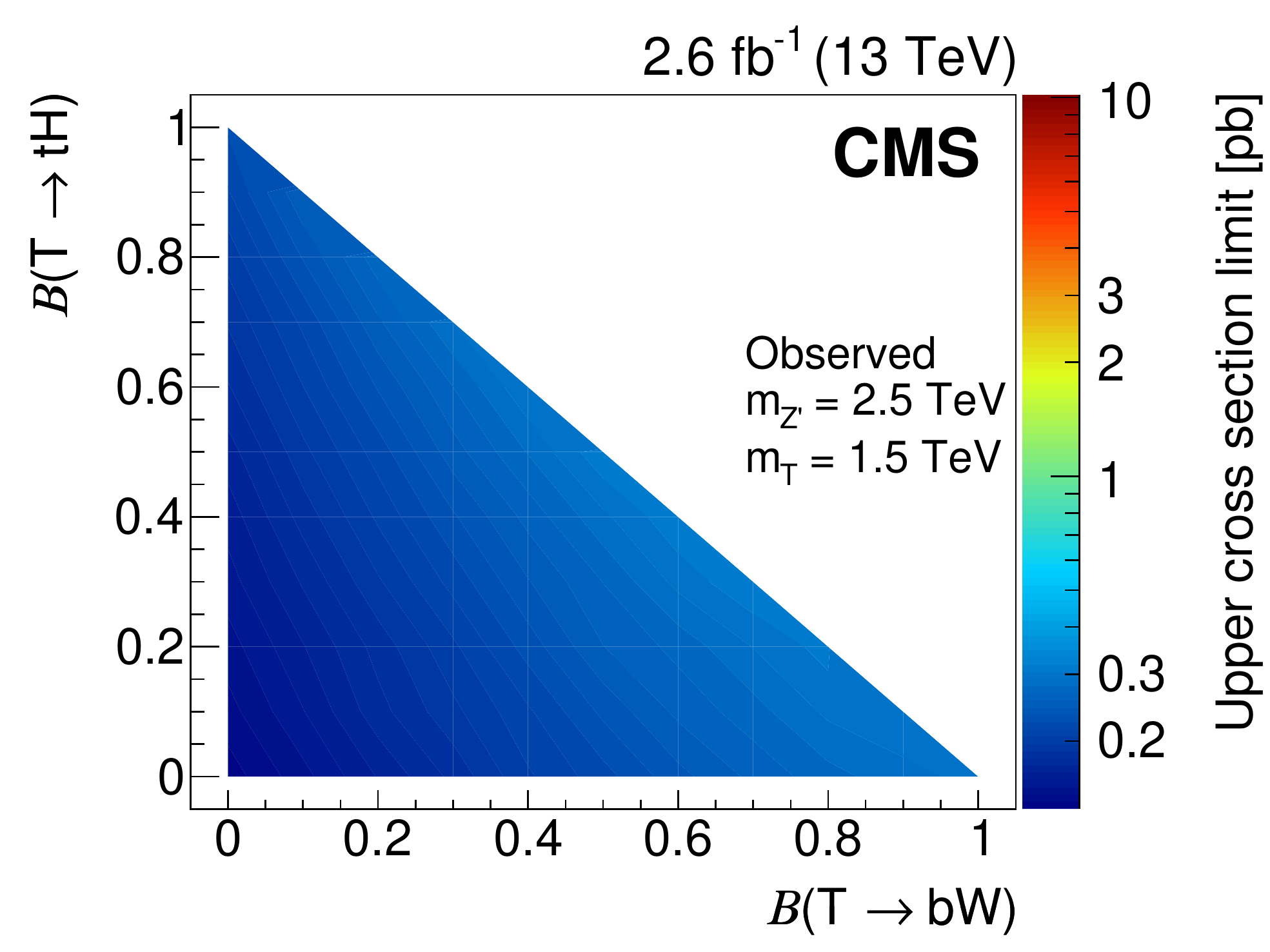}
\caption{Observed cross section limits for $\PZpr\to \PQT\PQt$ for different hypotheses for the $\PZpr$ boson and T quark masses, and the branching fraction of the T quark decay into bW and tH channels, with $\mathcal{B}(\PQT\to\PQt\Z)=(1-\mathcal{B}(\PQT\to \PQb\PW, \PQt\PH))$.
}
\label{obstriangle}
\end{figure}

Figures~\ref{exptriangle} and \ref{obstriangle} show the expected and observed upper cross section limits, respectively, for $\PZpr\to \PQT\PQt$ for different hypotheses of the $\PZpr$ boson and T quark masses, and the branching fraction of the T quark into bW and tH channels, with ${\mathcal{B}(\PQT\to\PQt\Z)}=(1-\mathcal{B}(\PQT\to \PQb\PW, \PQt\PH))$. Observed cross section limits are in all cases within 2 standard deviations of the expected values.

\begin{figure}[htbp]
\centering
\includegraphics[width=0.49\textwidth]{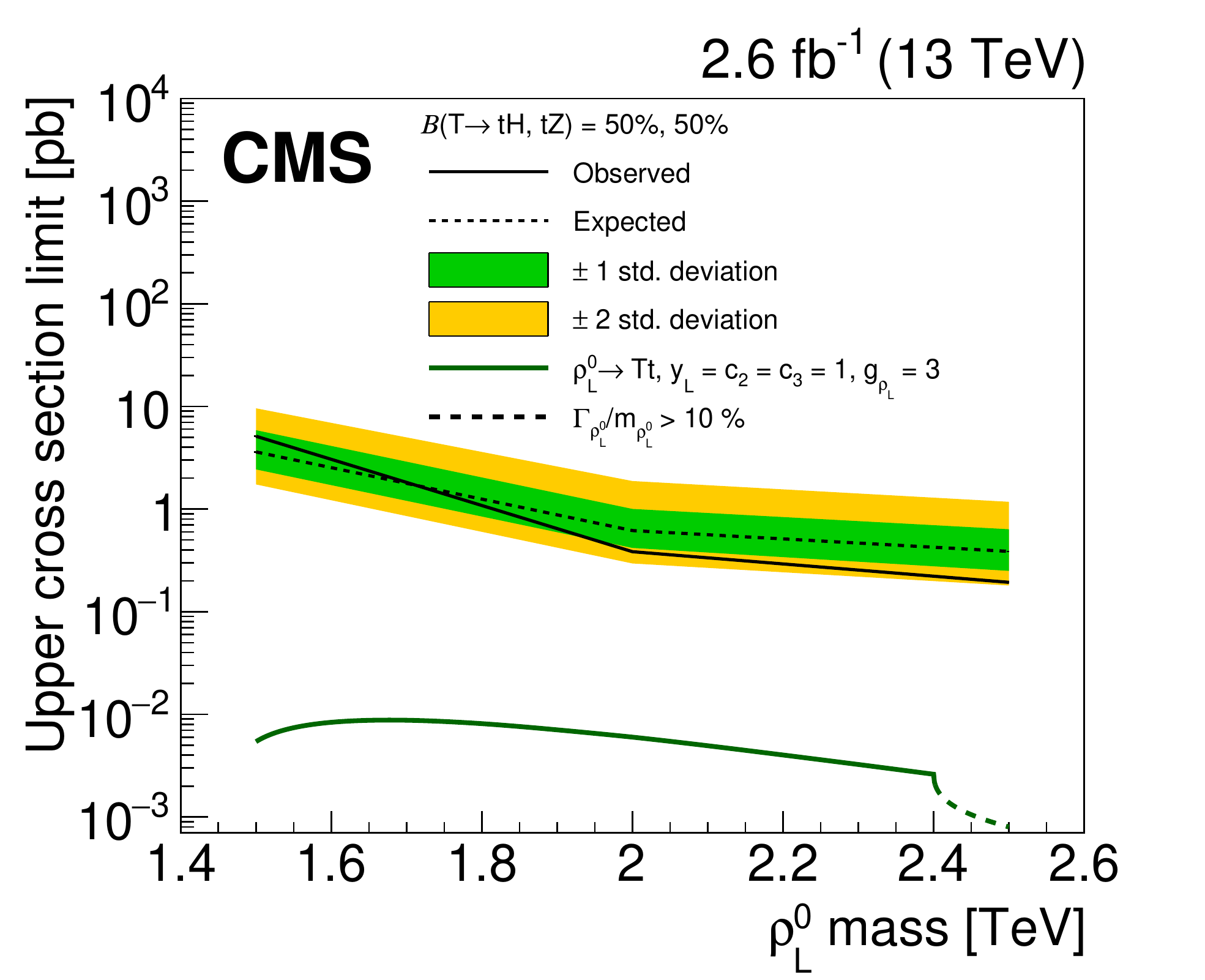}
\includegraphics[width=0.49\textwidth]{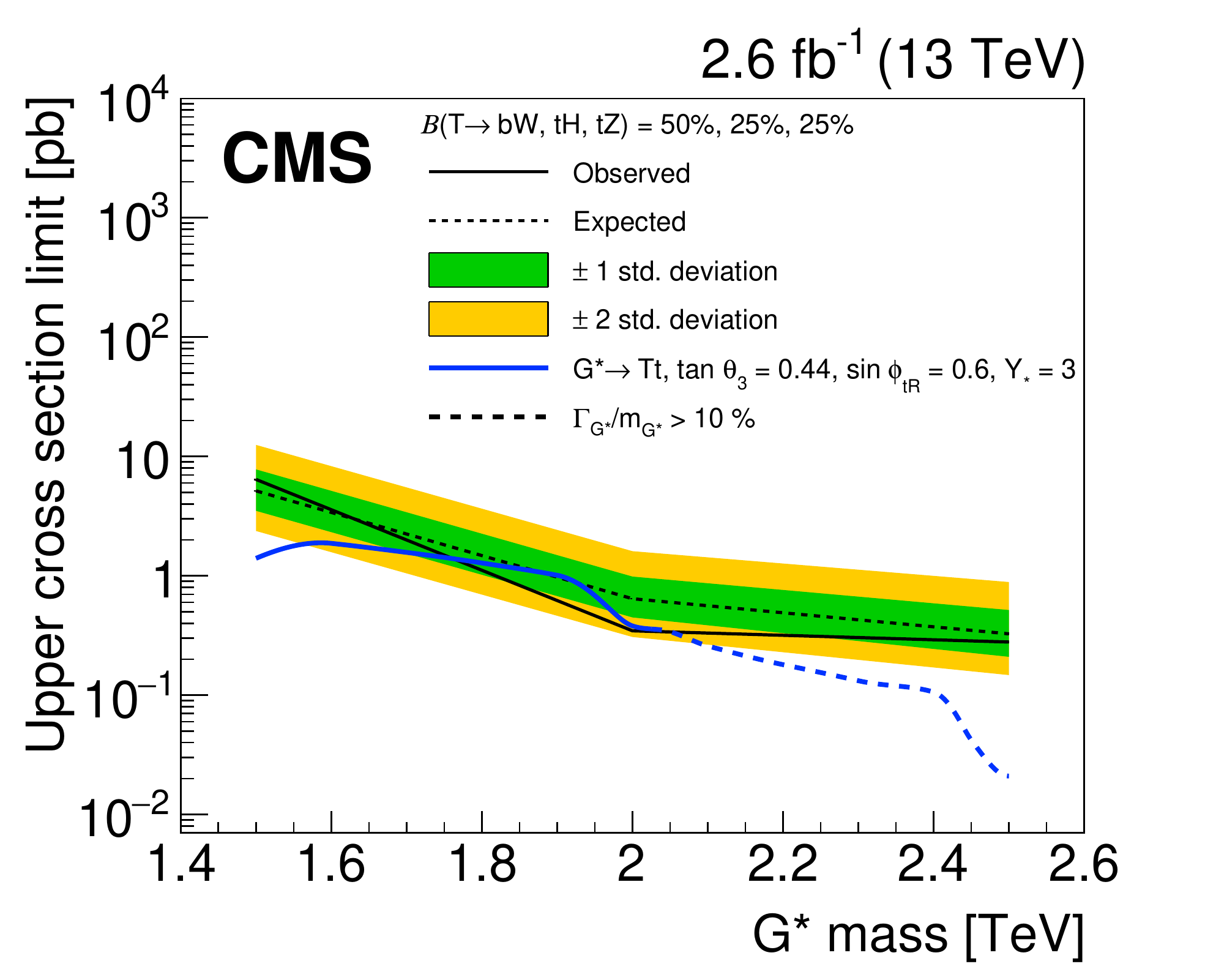}
\caption{One-dimensional cross section limits at 95\% CL as a function of the heavy vector resonance mass for $m_{\PQT}=1.2\TeV$, assuming branching fractions of the T quark decay to the tH/tZ channels of 50/50\% (left) or to the bW/tH/tZ channels of 50/25/25\% (right). The solid line is the observed limit, the dotted line is the expected limit, shown with 68\% (inner) and 95\% (outer) uncertainty bands. In the left plot, the green thick line is the product of the cross section and branching fraction for a heavy spin-1 resonance $\rho^0_L\to \PQT\PQt$ in a composite Higgs boson model \cite{Greco2014}. In the right plot, the blue thick line is the product of the cross section and branching fraction for a heavy gluon $\mathrm{G}^*\to \PQT\PQt$ in a warped extra-dimension model \cite{Bini2012}.  The theoretical predictions are shown as dashed lines where the width of the resonance is larger than 10\% of its mass.
}
\label{1dlim}
\end{figure}

One-dimensional cross section limits compared to the expectation of the composite Higgs boson model \cite{Greco2014}, as a function of the resonance mass for $m_{\PQT}=1.2\TeV$ and T branching fraction to tH/tZ channels of 50/50\%, are shown in Fig. \ref{1dlim} (left). A comparison of the limits to the warped-extra dimension model \cite{Bini2012} for T branching fractions to the bW/tH/tZ channels of 50/25/25\% is shown on the right-hand side of the same figure.
For some values of the mass of the heavy resonance, the resonance width is predicted to be larger than 10\% in the benchmark theoretical models. In these cases the simulated samples do not reproduce the behaviour of the theory benchmarks accurately, hence the cross section values are not considered for the comparison and are marked by a dashed line in Fig. \ref{1dlim}. The increase of the total width of the resonance with the increase of its mass is caused by additional decay channels becoming kinematically allowed. The change in slope of the theoretical cross sections around $m_{\PZpr}= 1.6$ and 2.4\TeV is due to the Tt and TT decay channels respectively becoming kinematically allowed. The comparison with the expectations of theoretical models shows that this search has no sensitivity to the composite Higgs model \cite{Greco2014} and some sensitivity to the extra dimensions model \cite{Bini2012}, however more data is needed to exclude specific scenarios.

\section{Summary}
\label{conclusions}
A search for a massive spin-1 resonance decaying to a top quark and a vector-like T quark has been performed in the all-jets channel using $\sqrt{s} = 13$\TeV proton-proton collision data collected by CMS at the LHC. The search uses jet-substructure techniques, involving top quark and W boson tagging algorithms, along with subjet $\cPqb$ tagging.
The top quark and W boson algorithms are based on the N-subjettiness variables and use the modified mass-drop algorithm to compute the jet mass.
The multijet background is estimated in data through a sideband region that is adjusted through simulation-based correction factors. The top quark background is estimated using  simulated events.

No excess is observed in data beyond the standard model expectations, and upper limits are set on the production cross sections of hypothetical signals.
The cross section limits are compared to the cross sections of a spin-1 resonance in a composite Higgs boson model and a Kaluza-Klein gluon in a warped extra-dimension model, for benchmark values of the model parameters, assuming a T quark mass of 1.2\TeV. Branching fractions of the T quark decay to the tH/tZ channels of 50/50\% and to the bW/tH/tZ channels of 50/25/25\% are assumed for models with a composite Higgs boson and with a warped extra-dimension, respectively. This search is not sensitive to the composite Higgs model \cite{Greco2014} with the analyzed data. In the case of the model with a warped extra-dimension \cite{Bini2012}, the upper limit obtained on the cross section is just at the predicted level for $\mathrm{G}^*$ masses in the region of 1.8\TeV.
Although limits are not placed on these particular models, more generally a $\PZpr$ boson decaying to a top and a T quark is excluded at 95\% confidence level, with upper limits on production cross sections ranging from 0.13 to 10\unit{pb}, depending on the hypotheses.
This is the first search for a heavy spin-1 resonance decaying to a vector-like T quark and a top quark.

\begin{acknowledgments}
We congratulate our colleagues in the CERN accelerator departments for the excellent performance of the LHC and thank the technical and administrative staffs at CERN and at other CMS institutes for their contributions to the success of the CMS effort. In addition, we gratefully acknowledge the computing centres and personnel of the Worldwide LHC Computing Grid for delivering so effectively the computing infrastructure essential to our analyses. Finally, we acknowledge the enduring support for the construction and operation of the LHC and the CMS detector provided by the following funding agencies: BMWFW and FWF (Austria); FNRS and FWO (Belgium); CNPq, CAPES, FAPERJ, and FAPESP (Brazil); MES (Bulgaria); CERN; CAS, MoST, and NSFC (China); COLCIENCIAS (Colombia); MSES and CSF (Croatia); RPF (Cyprus); SENESCYT (Ecuador); MoER, ERC IUT, and ERDF (Estonia); Academy of Finland, MEC, and HIP (Finland); CEA and CNRS/IN2P3 (France); BMBF, DFG, and HGF (Germany); GSRT (Greece); OTKA and NIH (Hungary); DAE and DST (India); IPM (Iran); SFI (Ireland); INFN (Italy); MSIP and NRF (Republic of Korea); LAS (Lithuania); MOE and UM (Malaysia); BUAP, CINVESTAV, CONACYT, LNS, SEP, and UASLP-FAI (Mexico); MBIE (New Zealand); PAEC (Pakistan); MSHE and NSC (Poland); FCT (Portugal); JINR (Dubna); MON, RosAtom, RAS, RFBR and RAEP (Russia); MESTD (Serbia); SEIDI, CPAN, PCTI and FEDER (Spain); Swiss Funding Agencies (Switzerland); MST (Taipei); ThEPCenter, IPST, STAR, and NSTDA (Thailand); TUBITAK and TAEK (Turkey); NASU and SFFR (Ukraine); STFC (United Kingdom); DOE and NSF (USA).

\hyphenation{Rachada-pisek} Individuals have received support from the Marie-Curie programme and the European Research Council and EPLANET (European Union); the Leventis Foundation; the A. P. Sloan Foundation; the Alexander von Humboldt Foundation; the Belgian Federal Science Policy Office; the Fonds pour la Formation \`a la Recherche dans l'Industrie et dans l'Agriculture (FRIA-Belgium); the Agentschap voor Innovatie door Wetenschap en Technologie (IWT-Belgium); the Ministry of Education, Youth and Sports (MEYS) of the Czech Republic; the Council of Science and Industrial Research, India; the HOMING PLUS programme of the Foundation for Polish Science, cofinanced from European Union, Regional Development Fund, the Mobility Plus programme of the Ministry of Science and Higher Education, the National Science Center (Poland), contracts Harmonia 2014/14/M/ST2/00428, Opus 2014/13/B/ST2/02543, 2014/15/B/ST2/03998, and 2015/19/B/ST2/02861, Sonata-bis 2012/07/E/ST2/01406; the National Priorities Research Program by Qatar National Research Fund; the Programa Clar\'in-COFUND del Principado de Asturias; the Thalis and Aristeia programmes cofinanced by EU-ESF and the Greek NSRF; the Rachadapisek Sompot Fund for Postdoctoral Fellowship, Chulalongkorn University and the Chulalongkorn Academic into Its 2nd Century Project Advancement Project (Thailand); and the Welch Foundation, contract C-1845.
\end{acknowledgments}

\bibliography{auto_generated}

\cleardoublepage \appendix\section{The CMS Collaboration \label{app:collab}}\begin{sloppypar}\hyphenpenalty=5000\widowpenalty=500\clubpenalty=5000\textbf{Yerevan Physics Institute,  Yerevan,  Armenia}\\*[0pt]
A.M.~Sirunyan, A.~Tumasyan
\vskip\cmsinstskip
\textbf{Institut f\"{u}r Hochenergiephysik,  Wien,  Austria}\\*[0pt]
W.~Adam, E.~Asilar, T.~Bergauer, J.~Brandstetter, E.~Brondolin, M.~Dragicevic, J.~Er\"{o}, M.~Flechl, M.~Friedl, R.~Fr\"{u}hwirth\cmsAuthorMark{1}, V.M.~Ghete, C.~Hartl, N.~H\"{o}rmann, J.~Hrubec, M.~Jeitler\cmsAuthorMark{1}, A.~K\"{o}nig, I.~Kr\"{a}tschmer, D.~Liko, T.~Matsushita, I.~Mikulec, D.~Rabady, N.~Rad, B.~Rahbaran, H.~Rohringer, J.~Schieck\cmsAuthorMark{1}, J.~Strauss, W.~Waltenberger, C.-E.~Wulz\cmsAuthorMark{1}
% ...
\vskip\cmsinstskip
% ...
1:~~Also at Vienna University of Technology, Vienna, Austria\\
% ...

\end{sloppypar}
\end{document}